\def\gtapprox{\mathrel{%
   \rlap{\raise 0.511ex \hbox{$>$}}{\lower 0.511ex \hbox{$\sim$}}}}
\def\ltapprox{\mathrel{
   \rlap{\raise 0.511ex \hbox{$<$}}{\lower 0.511ex \hbox{$\sim$}}}}
\def\nl{\hfil\break}
\magnification=1200
\hsize 6truein
\vsize 8.75truein
\hoffset .25truein
\tolerance=2000
\baselineskip=21pt
\headline={\hss\tenrm\folio\hss}
\footline={\hfil}

\def\ref{\hangindent 2pc \hangafter 1}
\def\nl{\hfil \break}
\parskip 10pt plus 1pt
\parindent 0pc
\vskip 0.8truein
\vskip 0.8truein
\vskip 0.8truein
\vskip 0.8truein
\centerline{\bf MULTIPLE MOLECULAR WINDS IN EVOLVED STARS I. A SURVEY}
\centerline{\bf OF CO(2-1) AND CO(3-2) EMISSION FROM 45 NEARBY AGB STARS}
\vskip 0.8truein
\centerline{G.R. $\rm Knapp^1$, K. $\rm Young^2$, E. $\rm Lee^1$, and
A. $\rm Jorissen^3$}
\centerline{gk@astro.princeton.edu}
\centerline{rtm@dolson.harvard.edu}
\centerline{erick@astro.princeton.edu}
\centerline{ajorisse@astro.ulb.ac.be}
\vskip 0.8truein
$\rm ^1Department$ of Astrophysical Sciences, Princeton University,
Princeton, NJ 08544\nl
$\rm ^2Harvard-Smithsonian$ Center for Astrophysics, 60 
Garden St., Cambridge, MA 02138\nl
$\rm ^3U$niversit\'e Libre de Bruxelles, C.P. 226, Boulevard de
Triomphe, B-1050 Bruxelles, Belgium\nl
\vfil
\eject
\parindent 2pc
\line{\bf Abstract\hfil}
\bigskip

This paper describes observations of a new phenomenon in evolved
mass-losing AGB stars: the presence of two winds with different
expansion velocities.  CO(2-1) and CO(3-2) line emission was observed for
45 AGB stars at high velocity resolution and double winds found
in 20\% of the sample.  Highly asymmetric lines were found in six
other stars.  The data tentatively suggest that double winds occur
when the star undergoes a change (pulsational mode, chemical composition)
and that the very narrow components represent the onset
of a new phase of mass loss.

\bigskip
\bigskip
\line{\bf 1. Introduction \hfil}

\bigskip

Highly evolved stars on the asymptotic giant branch (AGB)
and beyond are surrounded by extensive circumstellar envelopes 
produced by the copious mass loss which dominates the evolution
at this stage.  Observations of infrared continuum emission
from circumstellar dust and millimeter wavelength line emission 
from circumstellar molecules
give the mass loss rates, wind outflow speeds, chemistry and structure
of the winds and allow the mass loss history of the star to be traced.
Several extensive data compilations which
provide the empirical basis for investigating
the mass loss phenomenon have been made (e.g. Loup et al. 1993), and
recent reviews include those by Habing (1996) and Olofsson (1996, 1997a,b).

Circumstellar molecules emit both maser and thermal radiation, and the latter
is modeled with a fair degree of success by calculating the emission from
a spherically symmetric wind expanding at a uniform velocity at a 
uniform mass loss rate, with the effective envelope radius determined
by molecular photodissociation by the interstellar radiation field
(Morris 1980; Mamon, Glassgold and Huggins 1988; Kastner 1992; 
Groenewegen 1994).  This model predicts simple line shapes; optically
thick emission lines are parabolic while optically thin lines are
flat-topped, and
the full width of the line at zero power is twice the terminal outflow
speed of the wind.

While this model (hereafter called
the ``steady wind'') gives a decent
first order representation of most circumstellar shells, recent, more
detailed observations have found many shells which deviate from this
simple description:

\noindent
1. Bipolar outflows.  The shells around many mass-losing stars (e.g.
VY CMa, AFGL 2688, and OH231.8+4.2) show flattened, bipolar structure
on scales of $\rm \sim 10^{16}$ cm (e.g. Olofsson 1997a).

\noindent
2. Very fast molecular winds.  The highest steady wind outflow speeds, up
to $\rm \sim 40 ~ km~s^{-1}$, are found from a few supergiant stars
such as IRC+10420  and VY CMa (Loup et al. 1993; Knapp et al. 1997b).
A small number 
of stars have, as well as the steady wind component, molecular emission at
much higher speeds, in excess of
$\rm 200 ~ km~s^{-1}$ in the most extreme cases
(Cernicharo et al. 1989; Gammie et al. 1989; Young et al. 1992;
Jaminet et al. 1992; Alcolea et al. 1996; Knapp, Jorissen and
Young 1997; Sahai and Nyman 1997).  The
fast winds do not have parabolic line shapes, and it is unlikely that 
they are produced by constant-velocity outflow.  When observed with
sufficient angular resolution, fast winds appear to be bipolar
and to be flowing from the poles of dense disks or toroids (Neri
et al. 1992; Yamamura et al. 1994).

\noindent
3. Interrupted mass loss.  Olofsson et al. (1996) describe observations
of detached molecular shells around several carbon stars, which appear
to be produced on thermal pulse timescales.

\noindent
4. Multiple-velocity and/or non-spherical outflows.  Margulis et al.
(1990) pointed out that not all molecular line profiles from evolved 
stars are adequately described by the steady wind model; some lines 
appear to be roughly gaussian or triangular in shape.  Some few of
these non-parabolic profiles have since been observed with
sufficient angular and/or velocity resolution to demonstrate that they appear
to have two or more components, centered at
the stellar systemic velocity but with different outflow velocities.  Among
the stars showing this phenomenon are Mira (Planesas et al. 1990a,b), X Her 
(Kahane and Jura 1996) and RS Cnc (Jorissen and Knapp 1997). The profile
shapes differ from those seen for stars with fast winds in that both
components have roughly parabolic shapes and the outflow speeds lie
within the range (3 - 30 $\rm km~s^{-1}$) observed for the large majority 
of evolved stars.  These stars thus appear to have {\it two}
steady winds.

Recent improvements
in receiver sensitivity, frequency coverage, frequency resolution
and baseline stability make it feasible to carry out low noise,
high velocity resolution observations of a significant sample of 
circumstellar envelopes to investigate the frequency of occurrence
of the multiple wind phenomenon and how it is related to 
the stellar chemistry, mass loss rate and evolutionary stage. This paper begins
the discussion and analysis of a series of observations made to investigate
multiple slow molecular winds in circumstellar envelopes.

\bigskip

\line{\bf 2. Observations \hfil}

This paper describes observations of the CO(2-1) and CO(3-2) lines made
with the 10.4 m Robert B. Leighton telescope of the Caltech 
Submillimeter Observatory on Mauna Kea, Hawai`i.  The stars to be observed
were selected from large surveys of CO emission (Margulis et al. 1990;
Nyman et al. 1992; Loup et al. 1993; Knapp et al. 1997b) as having
strong CO line emission, i.e. brighter than about 1 K
as observed with a 10 m telescope.  We also confined the observations
to stars north of about $-30^o$ because of the significant atmospheric
opacity at submillimeter wavelengths.  We can therefore expect to make
observations with a high signal to noise ratio at high velocity
resolution in a reasonable amount of observing time (2 hours or less).
The observations were mostly carried out in the period
Dec 28 1996 to Jan 2 1997, when we observed almost all CO-bright stars
in the right ascension range available at that time of year, about
$\rm 23^h$ to $\rm 10^h$.  The observed sample also contains several stars
with weaker CO emission which were observed to fill empty places in the
schedule and a small number of observations of stars in other
parts of the sky
obtained in the previous two years, during test and engineering time.
The sample of
stars described in this paper is thus very incomplete in terms of sky coverage,
and we plan to complete these observations in future observing sessions.

In all, high velocity resolution and (mostly) high signal-to-noise ratio
observations were obtained for 43 stars, which are listed in Table 1.
Table 1 also contains data for
two additional stars, both of which are known from previous
observations to have double outflows: o Ceti (Planesas et al. 1990a,b; Knapp
et al. 1997a), and $\Pi^1$ Gru (Sahai 1992).  The CO(2-1) observations of RS Cnc
are from Jorissen and Knapp (1997).

Table 1 lists the star's most common name, the observed 
position, the spectral type, the variable type, the basic chemistry
(O = oxygen rich, C = carbon star, S = S star), the period where known
and a simple classification for the shape of the CO line profile (see
below).

Most of the spectral types and chemistry classes are from Loup et al. (1993)
or from the SIMBAD listings.  Most of the variable types and periods
are from the Catalogue of Variable Stars by Kholopov et al. (1985); the
period of $\rm 55^d$ listed in that catalogue for EP Aqr is uncertain.
The periods for several stars with thick circumstellar
shells (R Scl, R Lep, AFGL 865,
AFGL 971, IW Hya, IRC+10216 and IRC+40540) are given by Le Bertre (1992, 
1993) and Cohen and Hitchon (1996).  The period of the post-AGB carbon star
HD 56126 is from L\`ebre et al. (1996).  Several of the variable types in
column 5 are from Kerschbaum and Hron (1992, 1994); these authors show
that the semi-regular SRa class (Kholopov et al. 1985) contains a 
mixture of Mira and SRb (hereafter ``true'' semiregular variables) and
have reclassified several stars; R Scl from SRa to SRb and BK Vir
and $\Pi^1$ Gru from Lb (cool irregular variable, see Kerschbaum et al. 
1996) to SRb.

The molecular line observations were made in the CO(2--1) or CO(3--2) line
depending on the weather, which was variable during the observing run.
The observations were made using liquid-helium cooled SIS junction receivers
with double-sideband system temperatures of $\sim$ 100 K and 130 K
at 230 and 345 GHz.  The telescope half-power beamwidths are $\rm 30''$
and $20''$ at these frequencies, and the main-beam efficiencies 
are 76\% and 65\%.  The receivers for both frequency bands are double
sideband, and spectral lines in the image sideband, 3 GHz below the 
observed frequency for both receivers, were also detected.

The spectral lines were observed by three acousto-optic spectrographs (AOS),
with bandwidths of 1.5 GHz over 2048 channels, 500 MHz over 1024 channels
and 50 MHz at 1024 channels.  The velocity resolution of the 50 MHz AOS
is $\rm \sim 0.12 ~ km~s^{-1}$ at 345 GHz and $\rm \sim 0.2 ~ km~s^{-1}$
at 230 GHz.  The spectrometer frequency was calibrated using an 
internally generated frequency comb, and the velocity scale is corrected
to the Local Standard of Rest (LSR).

The observations were made by chopping between the star position and an 
adjacent sky position with the secondary mirror, using a  chop throw of
$\rm 120''$ in azimuth at a rate of 1 Hz.  Pairs of chopped observations
were made with the source placed alternately in each beam. The spectral
baselines resulting from this procedure are linear to within the r.m.s.
noise for all bandwidths with a small number of exceptions for which the 
500 MHz data show small (0.005  - 0.01 K) baseline non-linearities.
The CO emission from all of the observed stars
is strong enough that it was used to measure the telescope pointing
offsets before each observation.

The temperature scale and the atmospheric opacity were measured by 
comparison with a hot (room temperature) load.  The line temperature
was corrected for the main beam efficiency and the resulting scale is
the Rayleigh-Jeans equivalent main-beam brightness temperature, $\rm T_{MB}$,
i.e. that measured by a perfect 10.4 m antenna above the atmosphere.
The high velocity resolution CO line profiles are shown in Figure 1.

Most of the line profiles are closely approximated by parabolic 
or flattened parabolic shapes
(e.g. AFGL 865, W Aql).  These line profiles are classified
in Table 1 as parabolic, P.  The CO(2-1) line profile for IRC+10216
shows the small blue-shifted feature which can be attributed to radiative
transfer effects in a shell with a monotonically decreasing radial 
temperature gradient and a small amount of turbulence (Huggins
and Healy 1986).  Two of the
profiles, those for  AFGL 971 and V Cyg, are affected
by galactic emission and hence their profile shapes are uncertain.
There is strong galactic emission near 0 $\rm km~s^{-1}$ in the `on' and
`off' positions seen in the broad-band profile of IRC+60144 which may
affect the observed line shape.
All three line profiles are classified as parabolic.
The observations of several other
stars (R And, S CMi, S Vir, V CrB and R Ser) were made with insufficient 
signal to noise ratio to measure the profile shape, and are classified as
parabolic by default.  Uncertain classifications are enclosed in
parentheses.

Several line profiles are distinctly non-parabolic.  Most striking are
those showing two winds, i.e. two components of different width
centered at the same velocity; examples are IRC+50049, X Her and EP Aqr.
These are classified as `D', or double-wind profiles.  Several other stars,
e.g. R Scl and R Leo, have asymmetric line profiles, with one side of the
profile being brighter than the other.  These are classified as A (for 
asymmetric) Blue or Red, depending on whether the red or blue side of the
profile is stronger.  The possible origin of these line profile shapes
will be discussed elsewhere. The stars with asymmetric profiles will
be included in the group with parabolic line profiles in the discussion
below, where stars in this combined group are compared with those with
double winds.

In almost all cases of stars with double winds, the narrow component has
a similar width to the emission lines from interstellar molecular clouds
- compare the observations of the contaminated profile of AFGL 871 and
of the double-wind profile for EP Aqr in Figure 1.  Are these double
wind line profiles, then, simply an artifact of superimposed emission from
molecular clouds?  We believe not, for the following reasons.  First,
mapping observations of several of these stars: o Cet (Planesas et al. 1990a,
b); RS Cnc (Neri et al. 1997); X Her (Kahane and Jura 1996); $\rm \Pi^1$
Gru (Sahai 1992); and EP Aqr (Knapp et al. 1998) show that both the broad and
narrow components are co-located, centered on the star and of small angular
extent.  Second, the central velocities of both the broad and the narrow
components are closely the same (see Table 2 below).  We therefore have high
confidence that both the broad and narrow components observed in the 
emission profiles represent circumstellar gas produced by mass loss.

How well can the presence of double winds be found in these observations?
A series of simulated CO(2-1) line profiles, with two components centered on
the same velocity but with different outflow velocities, was constructed
to investigate this question (for details see Section 4c below). 
In all cases, the fast wind component 
has an outflow speed $\rm V_f ~ = ~ 10 ~ km~s^{-1}$ and a mass loss rate 
of $\rm 3 \times 10^{-7} ~ M_{\odot} ~ yr^{-1}$.  The model envelope
was assumed to be at a distance of 200 pc and to be observed with the
CSO 10.4 meter telescope; the resulting spectral line is parabolic
with a peak temperature of $\rm T_f$ =  0.65 K. The models were calculated
with a velocity resolution of 0.1 $\rm km~s^{-1}$, an r.m.s. noise of 0.4 K,
and flat baselines.

A second component was then added, with varying peak temperature $\rm
T_S$ and outflow speed $\rm V_s$.  These simulations showed that:

\noindent
(1) If $\rm V_s/V_f ~ > ~ 0.7$, the resulting line profile is sufficiently
parabolic in shape that the presence of two winds cannot be distinguished.

\noindent
(2) If $\rm 0.7 ~ < ~ V_s/V_f ~ < 0.5$, the presence of the two winds can
be seen in a high-quality line profile provided $\rm T_s ~ \sim ~ T_f$.

\noindent
(3) If $\rm V_s/V_f ~ < ~ 0.5$, the slow wind can always be distinguished
provided that $\rm T_s$ is greater than five times the r.m.s. noise.

\noindent

A selection of the simulated profiles is shown in Figure 2.

The bottom line is that we can only reliably detect a double wind
when there is a reasonable contrast between the fast and slow
winds, i.e. when the outflow velocities differ by a factor of two or more.
We therefore cannot use the data in this paper to rule out the 
possibility that {\it all} stars have multiple winds with similar
expansion velocities.  This caution applies to all of the discussion
and conclusions in the remainder of this paper.  It is of interest in this
context that high spatial resolution far-infared maps of two of the carbon
stars in Table 1, U Hya and Y CVn, find double/detached shells (Waters
et al. 1994; Izumiura et al. 1996) but the global CO line profiles 
for these stars (Figure 1) show no discernable velocity structure.

\bigskip
\line{\bf 3. Line Profiles and Stellar Properties\hfil}
\bigskip

\line{\bf a. Chemistry\hfil}

Table 1 lists 45 stars: 26 are oxygen-rich, six are S stars and 13 are carbon
stars.  Six of the stars have asymmetric profiles, and all of these save R
Scl are oxygen-rich.  Nine of the stars have double profiles: three of these 
are S stars, one is a carbon star and five are oxygen stars; however,
the presence of a narrow component for the carbon star IRC+60144
is somewhat uncertain (see Figure 1). The remainder
have parabolic line profiles (and a few are observed with insufficient
sensitivity to define the line shape well.)  Thus, at least 38\% of the 
oxygen stars, 50\% of the S stars and 15\% of the carbon stars have
non-parabolic line profiles.  These results show the trend noted by 
Margulis et al. (1990); the line profiles from oxygen-rich circumstellar
envelopes have more structure than do those for carbon-rich envelopes.

\bigskip
\line{\bf b. Pulsational Properties\hfil}

Figure 3 shows the histogram of the periods of stars with double and with 
parabolic line profiles; the former tend to have shorter periods.  This
difference is also apparent when variable type is considered; most (20)
of the 26 Mira variables have parabolic profiles, while only half of the 
15 SR variables do.  Further, while the period distribution of the
whole sample of stars has a peak at around 300 days, this peak is far
more pronounced in the stars with double winds.

\bigskip
\bigskip
\line{\bf c. Infrared Colors\hfil}

Figure 4 shows the [12] - [25], [25] - [60] and K - [60] infrared 
colors for the stars in Table 1, showing the dependence both on profile
shape and on chemistry.  Here, the colors are referred to those of a 
$\rm 0^m$ star, using the zero point flux densities in the IRAS
Explanatory Supplement (1988).  The 12$\rm \mu m$, 25$\rm \mu m$ and
60$\rm \mu m$ flux densities from which these colors are calculated are 
from the IRAS Point Source Catalogue (1988) except for the S stars,
whose flux densities are given by Jorissen and Knapp (1997). The
K band magnitudes are taken from the compilation by Gezari et al.
(1993), from Le Bertre (1992, 1993) and from Kerschbaum and Hron (1994).
There may be large uncertainties in these colors, especially in K - [60],
because the flux densities are not observed at the same point in the star's 
variability cycle (typically, $\rm \Delta K ~ \sim ~ 1^m$, Le Bertre
1992).  Figures 4a and 4c show [25] - [60] vs [12] - [25] (cf. van
der Veen and Habing 1988) and K - [60] versus [12] - [25], respectively,
showing the colors as a function of profile shape, while Figures 4b and 4d
show these relationships as a function of envelope chemistry. Generally,
the higher the mass loss rate the redder the infrared colors
(van der Veen and Habing 1988).  Figure 4 shows that stars with double winds
are not present in the regions of these color-color diagrams occupied by
stars with high mass loss rates;  that this is not simply an artifact 
of the dependence of infrared color on circumstellar chemistry
is shown by comparison with Figures 4b and 4d.

\bigskip
\line{\bf d. Summary\hfil}

Nine of the 45 evolved stars in the sample in Table 1 have double winds, and
only one of these is a carbon star.  The stars with double winds have infrared
colors which suggest that they have mass loss rates lying at the low end
of those in the sample, and also have shorter periods than average.
Further, three of the six S stars have double winds.  There are two
tentative conclusions from these properties: (1) it is unlikely that
the double wind phenomenon is the result of binarity, since the frequency
of double winds appears to depend on stellar chemistry and variable type, both
of which are intrinsic to the star: and (2) they suggest that the 
double-wind phenomenon may be associated with period irregularities
and occur when the chemistry and perhaps stellar pulsation and mass loss 
are undergoing a transition from one phase to another.

\bigskip
\line{\bf 4. Physical Properties of the Circumstellar Envelopes\hfil}
\bigskip

\line{\bf a. Kinematics\hfil}

The parameters of the spectral lines, measured from the high velocity
resolution profiles, are given in Table 2, which lists for each star the
CO transition observed, the  r.m.s. noise, the integrated CO line intensity
$\rm I_{CO}$ in $\rm K \times km~s^{-1}$, and the peak line temperature 
$\rm T_{peak}$, the central velocity $\rm V_c$ and the terminal wind outflow
velocity $\rm V_o$.  The last three quantities were calculated for most
of the line profiles by fitting a flattened parabolic line profile of the 
form
$$\rm T(V) ~~ = T_{peak} ( 1 - exp[-{\alpha}
(1  - ({{V - V_c}\over
{V_o}})^2)]) / ( 1 - e^{-\alpha}) ~~~ \eqno(1)$$
(Knapp and Morris 1985) to the data.  This was also done for 
the asymmetric line profiles such as that for R Leo to yield average
quantities.  The line profiles for IRC+60144, AFGL 971 and V Cyg
are partly contaminated by galactic emission.  Equation 1 was fit to 
the uncontaminated regions of the profile, and the 
integrated line intensity $\rm I_{CO}$ was estimated assuming that 
the line shape is parabolic: $\rm I_{CO} ~ = ~ 4/3 T_{peak} V_o$.
These values are enclosed in parentheses in Table 2.

The line profiles for stars with two components were modeled by a 
superposition of two parabolas.  No fit was made for the CO(2-1)
profile for R Hya; this profile is very asymmetric.  CO(2-1)
data for RS Cnc from Jorissen and Knapp (1997) and CO(3-2) data
for o Cet from Knapp et al. (1997a) were also analyzed.  The profile 
fits for the stars with double winds are shown in Figure 5.  Table 2 
lists the parameters for the broad and narrow components separately.
As Figure 5 shows, these fits are adequate for most of the stars.
The narrow components for several stars appear to be more gaussian than
parabolic in shape.  Two stars, o Cet and R Hya, have broad 
components whose redshifted emission is stronger than the blueshifted
emission; the 
profiles for the remaining stars are roughly symmetric.

The central velocities $\rm V_c$ and outflow velocities $\rm V_o$
measured by the CO(2-1) and CO(3-2) lines agree well for the stars
for which both lines have been observed.  The central velocities
of the broad and narrow components are the same to within the
errors for all stars, with $\rm <V_c(broad) ~ - ~ V_c(narrow)>
~ = -0.2 ~ km~s^{-1}$.  Thus both flows are centered on the star
(and among other things this means that it is very unlikely that
the narrow components are due to a chance superposition of interstellar
emission on the stellar line profile).  Figure 6 shows the 
histograms of the outflow velocities for stars with double winds:
the outflow velocities of the broader component are shown in 
Figure 6a and of the narrower component in Figure 6b.  For comparison,
the histogram of outflow speeds for the stars in Table 2 with
single winds is shown in Figure 6c.  Figure 6c appears to contain 
data for a representative sample of stars; the distribution
is very similar to that found for other samples
(see, e.g. Zuckerman, Dyck and Claussen 1986).  
The distributions
of outflow velocities in stars with double winds differ from those in 
stars with single winds (although this difference is subject to the 
difficulties in distinguishing two components when the outflow velocities
have similar values - see the discussion in Section 2). 
The broad components have somewhat larger 
outflow velocities than typical, but much more striking is the difference
between Figures 6b and 6c; the narrow components have much smaller
outflow velocities than average.  Figure 6 suggests that the really 
unusual property of most of the stars with double winds is the presence of a
component with a very small outflow speed. (The exceptions are IRC+60144
with $\rm V_s ~ = ~ 5.6 ~ km~s^{-1}$ and $\rm V_f ~ = ~ 20.5 ~ km~s^{-1}$,
and $\Pi^1$ Gru, with $\rm V_S ~ = ~ 12 ~ km~s^{-1}$ and $\rm V_f ~ = ~
27 ~ km~s^{-1}$.) Indeed, Figure 6 suggests
that there are essentially {\it no} stars with single winds with outflow
speeds less than about 3 $\rm km~s^{-1}$, and that these low velocity
flows appear only in stars with double winds.

Figure 6c also shows the distributions of the wind speeds in stars with
double winds, i.e. the sum of the data from Figures 6a and 6b.  The
distributions for stars with single winds and with double winds are
quite different.  This implies that there is indeed an intrinsic difference
between the envelopes of stars whose line profiles show one component 
and those which show two components.

Figure 7 shows the velocity of the slow wind versus that of the fast
wind for the nine stars with double winds.  With the exception of
EP Aqr, the slow and fast wind velocities are proportional to each
other.
\bigskip

\line{\bf b. Distances \hfil}

In the next section, we model the circumstellar CO emission to 
derive values for the mass loss rates.  A prerequisite for this
modeling is a measurement or estimate of the stellar distance,
and this situation has changed dramatically in recent months
with the publication of the Hipparcos catalogue (Perryman et 
al. 1997 - see also van Leeuwen et al. 1997).  Thirty-three of the stars
in Table 1 are in the Hipparcos Main Catalogue, and 23 of these have
parallaxes measured at greater than the 2$\sigma$ level.  The
distances to these stars are taken directly from the Hipparcos
catalogue.  

Other methods for estimating distances to AGB stars include using a 
period - $\rm M_V$ or period - $\rm M_K$ relationship; assuming
a constant absolute K magnitude; or assuming a constant bolometric
luminosity (see, e.g. Claussen et al. 1987; Jura 1988; Young 1995;
Knapp et al. 1997b).  These methods can be re-evaluated using 
the Hipparcos distances, and a preliminary discussion is given below.

Several studies (e.g. Claussen et al. 1987, Jura 1988,
Jorissen and Knapp 1997) have assumed the value of $\rm M_K  =
-8.1^m$ found for carbon stars in the Magellanic Clouds (Frogel
et al. 1980).  The values of $\rm M_K$ for the stars in Table 1
with parallaxes, regardless of chemistry, are shown in Figure 8a.  These
were calculated from the K magnitudes in the compilation of Gezari
et al. (1993) and parallaxes from the Hipparcos catalogue. Apart from
the supergiant $\alpha$ Ori ($\rm M_K ~ = ~ -9.6^m$) the stars appear
to have a mean absolute K magnitude about $\rm 1^m$ fainter than
$\rm -8.1^m$.  However, all of these stars are losing mass 
and so have significant circumstellar extinction.  The
approximate K extinction was found from:
$$\rm A_K ~ = ~ 3 \times 10^7 ~~ \mathaccent 95 M (M_{\odot} yr^{-1})
~~ V_o^{-1} (km~s^{-1}) ~~ + ~~ 10^{-4} D (pc)  \eqno(2)$$
The first term is the circumstellar extinction and the second the 
interstellar extinction (with D the distance in pc).  These quantities
were found by assuming a gas to extinction ratio $\rm N_H/A_V
~ = ~ 2 \times 10^{21} ~ cm^{-2} ~ mag^{-1}$ for both the circumstellar 
and interstellar gas, and an extinction curve with R  = $\rm A_V/E_{B-V}$ 
= 3.1, which gives $\rm A_V/A_K ~ = ~ 8.9$.  The circumstellar extinction
was calculated assuming constant mass loss rate and outflow velocity, 
so that $\rm n(r) ~ \sim ~ r^{-2}$, and an inner radius for the dust
shell of $\rm 5 \times 10^{13}$ cm.  The extinction corrections for
the stars in Figure 8a were calculated from the mass loss rates given
by the present CO observations (see below) and the resulting distribution
of $\rm M_K$ is shown in Figure 8b.  It should be emphasized that this
calculation is very uncertain; the differences between circumstellar
and interstellar matter, and between envelopes of different chemical
composition, have not been considered; there is growing evidence (see, 
e.g., recent papers by Skinner et al. 1997 and Whitelock et al. 1997)
that AGB mass loss is anisotropic; and the apparent K magnitudes
are uncertain because of variability
(e.g. Le Bertre 1992).  It is clear from Figure 8b that the extinction
has been much overestimated for several stars, but the results do support
the use of a single absolute K magnitude, consistent with the Magellanic
Cloud  carbon-star
value of $\rm -8.1^m$, for all stars, and show that circumstellar
extinction is often significant.

For stars in the present study which do not have Hipparcos parallaxes,
the distances were found using one or more of the above methods.
Stars whose distances were found using $\rm M_K ~ = ~ -8.1^m$
were modeled in an iterative process; a first distance estimate was
made, the mass loss rate was calculated, the circumstellar extinction
was calculated, and the distance re-estimated.  In all cases, the 
distance estimates were checked to be sure that they are not discrepant
with the lower limits given by Hipparcos.  These distance estimates
are preliminary, and should be refined using a larger sample of stars;
the absolute magnitudes of carbon and oxygen stars should also be
further investigated.
Distances were also estimated in some cases using $\rm L_{bol}
= 10^4 ~ L_{\odot}$.  The only serious inconsistency was found for R
Hya, whose ``K magnitude'' distance (115 pc) is significantly smaller 
that the limit of $>$ 200-300 pc from the Hipparcos parallax - see
also van Leeuwen et al. 1997b).  No distance estimate was made for the 
post-AGB star HD 56126.
\bigskip

\line{\bf c. Mass Loss Rates \hfil}

The mass loss rates were calculated using the model of Morris (1980) and 
Knapp and Morris (1985) to calculate the emission from a model spherical
envelope produced by constant mass loss at constant outflow speed.
The envelope radius was assumed to be truncated by photodissociation of CO,
using the rates given by Mamon, Glassgold and Huggins (1988).  The 
$\rm CO/H_2$ abundance was taken as $\rm 5 \times 10^{-4}$ for
oxygen stars, $\rm 6.5 \times 10^{-4}$ for S stars and $\rm 10^{-3}$
for carbon stars (see Knapp et al. 1997b).  The slow and fast components for
stars with double winds were treated separately.  The results are
given in Table 3,
which gives for each star the distance, the chemistry, the assumed outflow
velocity $\rm V_o$, the resulting mass loss rate 
$\rm \mathaccent 95 M$ and the source of the 
distance.  These mass loss rates generally agree reasonably well with
previous estimates - that for IRC+10216, for example, is about a factor of
2 lower than the value found by Crosas and Menten (1997) after the different
model abundances are taken into account.

Figure 9 shows histograms of the mass loss rates for the separate double-wind
components: the broad component (Figure 9a) and the narrow component
(Figure 9b).  For comparison, the distribution for the stars in Table 3
with single winds is shown in Figure 9c.  The broad components have a
distribution similar to that of the stars with single winds; the narrow
components have mass loss rates which are much smaller than typical.

Several authors (e.g. Netzer and Elitzur 1993; Young 1995) have demonstrated
an empirical 
relationship between mass loss rate $\rm \mathaccent 95 M$ and outflow speed
$\rm V_o$ for winds from AGB stars which
provides a useful method for displaying and comparing data sets.
Such a relationship is expected for radiationally-driven dusty winds, 
since the grain-gas coupling becomes less effective for lower mass loss
rates (Gilman 1972; Goldreich and Scoville 1976; Netzer and Elitzur 
1993; Ivezic and Elitzur 1995; Crosas and Menten 1997).

The mass loss rate and outflow speed are plotted 
against each other in Figures 10 - 12; these figures show the same data
subdivided in several different ways.  Figure 10 shows the plot for stars 
with distances measured by Hipparcos and those measured by other means.
Both samples show a strong correspondence between $\rm \mathaccent 95 M$
and $\rm V_o$, but Figure 10 raises the interesting possibility 
that the distances to the stars found to have the highest mass loss rates
(most, though not all, are carbon stars) have been overestimated and hence
the mass loss rates have been overestimated:  these stars, with $\rm
\mathaccent 95 M ~ \geq ~ 3 \times 10^{-6} ~ M_{\odot}$, do not have 
Hipparcos parallaxes (they are optically too faint) and lie above
the relationship found for the stars which do.  This should be further
investigated with a larger sample.  The stars with Hipparcos distances
show $\rm \mathaccent 95 M ~ \sim ~ V_o^2$, a shallower relationship
than found by Young (1995).

Figure 11 shows the data broken down into single winds, the slow
components of double winds and the fast components. All types of
component appear to follow the same relationship between $\rm
\mathaccent 95 M$ and $\rm V_o$.  Figure 12 shows the data divided
by stellar chemistry; all of the stars appear to follow the same relationship.

\bigskip

\line{\bf 5. Discussion\hfil}
\bigskip

The results presented in the previous sections of this paper show that the 
phenomenon of evolved stars with double winds is quite common: at least
20\% of the stars observed with high velocity resolution have two
components centered at the same velocity.  The present data show that the
phenomenon is particularly common in S stars but this remains to be
confirmed by observations of a larger sample.  In addition, several stars
were found to have strongly asymmetric profiles.  These phenomena are
almost absent in carbon stars.  The tendency of this sort of velocity
structure to be found in semiregular variable stars, or stars with short 
periods, as well as in S stars, suggests that double winds accompany 
changes in the stellar properties (pulsation mode, period etc)
(see also Olofsson 1997a; Jorissen and Knapp 1997).

If all stars go through one or more stages in which winds at two 
different velocities are produced, the present data show that
they spend at least 20\% of their 
AGB evolution in such a phase.  The total lifetime of stars in the 
mass-losing phase on the AGB has been estimated at $\rm \sim ~ 2
~ \times ~ 10^5$ years (e.g. Young et al. 1993a,b), so that stars 
would spend $\rm \sim 4 \times 10^4$ years in a phase or phases
in which two winds are produced.

Studies of Mira variables have found a correlation between pulsation period
and mass loss rate (e.g. Jura 1988; van der Veen and Habing 1988), 
and between period and wind outflow speed (Dickinson and Chaisson
1973), demonstrating the role of pulsation in mass loss.  Semi-regular
variables, which generally have shorter pulsation periods, do not show
a period-velocity correlation (Dickinson and Dinger 1982).
Figures 13 and 14 show these relationships for stars in the present sample
whose periods are known (Table 1), sorted by variable type.  The period -
mass loss rate correlation for Mira variables in Figure 13 and the
period - outflow velocity correlation for Mira variables in Figure 14
are fairly well defined, but are not followed by the semi-regular 
variables, as found by Dickinson and Dinger (1982).  The data for Miras
with double winds shown in these figures suggest that the considerable
scatter in these relationships may be intrinsic, and that the stellar
properties (luminosity, period etc.) do not uniquely determine the mass loss
rate.

As the data discussed in this paper have shown, it is the slow winds
which are unusual; they have outflow speeds and mass loss rates far lower than
typical for evolved stars.  Further, as Figure 7, 13 and 14 show, both
the slow and fast winds are related to each other and to the stellar
properties.
Most of the stars with double winds have fairly low
mass loss rates.  As a result, the CO-emitting extent of the circumstellar
envelope, determined by photodissociation, is small.  Table 4 lists the
photodissociation radii $\rm R_p$ for both the slow and fast winds
for the stars with double winds; these were calculated using the 
results of Mamon, Glassgold and Huggins (1988) as the radii at which
the relative CO abundance has decreased to half of its original value.
Table 4 also gives the crossing time T = $\rm R_p/V_o$ for both fast and
slow winds.  If the double-line profiles are due to the cessation of mass 
loss at one outflow velocity/mass loss rate and its resumption at
a different outflow velocity/mass loss rate, these times are upper 
limits to the time interval which can elapse between the cessation of
mass loss and its resumption.  For several stars (e.g. RS Cnc) this
interval is $<$ 1000 years, far shorter than the approximately 5000
years observed for some carbon stars with detached circumstellar 
envelopes (e.g. Olofsson et al. 1996; Lindqvist et al. 1996).

However, the correspondence between the presence of discernable double winds
and stellar period, variability, type and mass loss rate suggest that the
phenomenon only occurs at particular stages in a star's evolution. By
analogy with carbon stars with detached envelopes, where the observations
show that mass loss has resumed at a much lower mass loss rate and outflow
speed (Olofsson et al. 1996; Lindqvist et al. 1996), and because
winds with very low velocities ($\rm V_o \leq 3 ~ km~s^{-1}$) seem
to occur only in stars with double winds, we tentatively conclude that
oxygen and S stars also undergo episodes of interrupted mass loss, and
that the slow winds represent its resumption.

The present observations add to the growing body of data which show
that the kinematics of circumstellar
envelopes are a lot more complex that they appear at first, that changes in
the stellar variability modes may produce abrupt changes in the mass loss 
rate, and that these changes are fairly common.  Future papers in this
series will discuss further survey observations, the systematics of slow and
fast winds, asymmetric profiles, the relationship to the detached shells
seen in some carbon stars (Olofsson et al. 1996) and individual stars in detail.

\bigskip
\line{\bf 6. Conclusions \hfil}

The CO(2-1) or CO(3-2) line profiles from 43 nearby AGB stars with bright
CO emission were measured with high velocity resolution and 
sensitivity, and combined with observations of two more such stars from the
literature.  We found:

\parindent=0pc

1. At least a third of the line profiles are strongly non-parabolic
in shape.  Six stars have asymmetric lines with bright spikes at the
red or blue extrema of the profile and nine have double profiles,
i.e. two
roughly parabolic lines of different width centered at the same velocity.

2. The stars with non-parabolic profiles have lower pulsation periods,
bluer infrared colors (indicating lower mass loss rates) and are
much more likely to be semiregular variables, than most of the stars
in the sample.  Only two of the carbon stars (15\%) have non-parabolic
profiles, half of the S stars do, and 40\% of the oxygen stars.  These
results suggest that complex winds, whose presence is shown by complex
line profiles, occur when the star is undergoing a change
in luminosity, pulsation mode or chemistry.

3. The broad and narrow components of the stars with double winds, analyzed
separately, follow the same mass-loss-rate/period/outflow velocity
relationships as do the stars with parabolic profiles.  The broad
components have mass loss rates and outflow velocities typical of mass
losing AGB stars, but the narrow components have far lower velocities and
mass loss rates than are seen in any other sample of evolved mass-losing
AGB stars. The narrow components may be due to the
resumption of mass loss after it has been stopped by some 
change in the stellar properties.

4. Comparison of the line widths of $\rm ^{29}SiO(8-7)$ (for oxygen stars)
or CS(7-6) (for carbon stars) with the CO(3-2) line widths
shows that the SiO/CS lines are systematically narrower than the CO line
for stars with $\rm V_o ~ \leq ~ 10 ~ km~s^{-1}$; above this velocity
the line widths are roughly equal.  These data suggest that the winds in stars
with low mass loss rates do not reach their terminal velocities
until relatively large distances, several $\times 10^{15}$ cm.

5. A search for very broad line wings in this sample of stars, 
like those associated
with post AGB stars (for example AFGL 618),
proved negative except (tentatively) in two stars, where it
need further observations.  The phenomenon of fast winds appears to be
confined to stars which are evolving
away from the AGB.  

\bigskip
\parindent=2pc

\line{\bf Acknowledgements \hfil}

It was a privilege to use the Hipparcos data base, and we would
like to express our appreciation to the Hipparcos team for making it
so openly available.
We thank the staff at CSO, especially Antony Schinkel, Maren Purves and
Tom Phillips, for their support of the observations, and the referee,
Kay Justtanont, for a prompt and helpful reading of the paper.  Astronomical 
research at the CSO is supported by the National Science Foundation
via grant AST96-15025.  
This research made use of the SIMBAD
data base, operated at CDS, Strasbourg, France.  Support for this work from
Princeton University and from the NSF via grant AST96-18503
is gratefully acknowledged.

\bigskip
\bigskip
\line{\bf Appendix A.  Search for Fast ($\rm V_o \geq 50 ~ km~s^{-1}$)
Molecular Winds\hfil}

A small number of evolved stars have very fast ($\rm V_o \geq 50 ~ km~ s^{-1}$)
molecular winds in addition to the slow ($\rm V_o \leq 20 ~ km~s^{-1}$)
`steady' winds.  These winds, some with speeds in excess of 200 $\rm 
km~s^{-1}$, have, to date, been found in stars which are evolving away from
the AGB.  There is evidence that the formation of fast winds occurs after 
that of slow winds, and that these fast winds are bipolar (Morris et al.
1987; Cernicharo et al. 1989; Gammie et al. 1989; Bachiller et al. 1991;
Neri et al. 1992; Young et al. 1992; Jaminet et al. 1992; Bujarrabal
et al. 1994; Alcolea et al. 1996; Knapp, Jorissen and Young 1997).  

The observations described in the present paper are sensitive enough,
and almost always
have sufficiently flat baselines, that the line profiles measured
with the 500 MHz AOS provide useful information on the presence or
absence of very fast winds.  As a guide,
CSO observations of AFGL 618 find that the ratio of the line brightness
temperatures of the fast and `steady' winds is $\sim$0.04 for the CO(2-1)
line and 0.09 for the CO(3-2) line (Gammie et al. 1989), where the 
temperature of the fast wind emission is measured at $\rm \pm 50
~ km~s^{-1}$ from the central velocity. (The 
difference in these ratios is probably due to the fast wind being
confined to the inner regions of the CRL 618 circumstellar envelope while
the envelope produced by the steady wind is more extended and is partly
resolved).  CSO observations of AFGL 2688 give ratios of about
$\sim$0.12 and 0.24 (Young et al. 1992), and of V Hya $\sim$0.07
and 0.06 (Knapp, Jorissen and Young 1997).

However, a word of caution is in order here - the observations in this
paper give information only on the fast wind {\it emission} from evolved
stars.  Sahai and Nyman (1997) describe their beautiful observations
of the ``Boomerang'' bipolar nebula in the CO(1--0) line where a fast wind
is seen in absorption against the microwave background; the wind is presumed
to be cooled by rapid adiabatic expansion to a temperature $<$ 3 K. Gas at
this temperature has negligible excitation of the CO rotational lines
so is not detectable in absorption or emission in any line except
CO(1--0).

The 500 MHz filter bank observations of the stars listed in Table 1 
were reduced and calibrated as described in Section 2, and the reduced
profiles were binned to a resolution of $\rm \sim 7 ~ km~s^{-1}$
and examined to see if high-velocity emission was present. The
results are given in Table 5. The only two
stars with any hint of such emission are R Lep and RX Boo, whose CO
line profiles are shown in Figure 15; these stars should be re-observed,
since this structure could be due to weak emission lines from other
molecules in the circumstellar envelope.  The CO(2-1) profile of
IRC+10216 has weak high-velocity wings, with T(wing)/T(peak)
$\sim$ 0.0048 - see Table 5.  However, this feature is almost certainly
instrumental; the frequency calibration comb, as measured by the AOS,
shows wings at about this level (Table 5).  This represents the real 
limit to the measurable brightness of broad-wing emission.
 
Otherwise, no high-velocity emission is present for any of the other
stars.  Table 5 lists the 5$\sigma$ upper limit on the brightness
temperature of fast wind emission and its ratio to the peak temperature
of the steady wind, also taken from the binned profiles.  Broad-band
data are not available for a small number of stars because the baselines
are non-linear. The ratios in Table 5 are
less than typically observed in stars with fast winds (see above) for the
large majority of the stars.  Almost all of these stars have late
spectral types, showing that they are still on the AGB.  The single exception
is HD 56126, whose spectral type is F5Iab and which is probably a
post-AGB star (van Winckel et al. 1996; Justtanont et al. 1996; van
Winckel 1997).  These data provide additional evidence (subject to the possible
existence of cold, fast winds discussed above) that the stage 
when fast molecular winds are produced is short-lived, and occurs at
an evolutionary stage beyond the AGB.
\vfil
\eject

\line{\bf Appendix B.  Comparison of SiO/CS and CO Outflow Velocities\hfil}

The broad band (500 MHz) CO(3-2) observations detected emission from the
CS(7-6) line from carbon stars or the $\rm ^{29}SiO(8-7)$ line from oxygen
stars in the image sideband: the rest frequencies of these lines are
342.883 GHz and 342.979 GHz respectively.  Five of the carbon stars
and seven of the oxygen stars have emission of sufficient strength in these
lines to allow a comparison of the line widths.  The outflow speeds $\rm
V_o$ were calculated by fitting a parabolic line to the CS/SiO and CO
line profiles.  Note that at the approximately 1 $\rm km~s^{-1}$ resolution
of the 500 MHz AOS, any structure in the line profiles is not resolved
and the fitted outflow velocities are likely to find an average of the slow and
fast wind velocities.

The results are listed in Table 6, which gives the values for $\rm V_o$
for the CO(3-2) and $\rm ^{29}SiO$ or CS lines calculated from the 500 MHz
line profiles, plus the profile classifications and the CO outflow
velocities from the 50 MHz observations (see Section 4). 

Comparison of the CO velocities measured with the 500 MHz and the 50 MHz
AOS spectrometers shows the effect of velocity resolution; the lines
are broadened by about $\rm 1 ~ km~s^{-1}$.  However, the high resolution
CO lines are broader than the CS/$\rm ^{29}SiO$ lines, showing
that the velocity differences in Columns 2 and 3 of Table 6 are real.  Three
of the stars in Table 6, RS Cnc, R Hya and X Her, have double winds; the data
tentatively suggest that both the slow and fast winds contribute to
the $\rm ^{29}SiO$ emission for RS Cnc and X Her, and that the $\rm ^{29}SiO$
emission arises only from the slow wind in R Hya.  This question should be 
re-examined with higher velocity resolution observations of the CS and
SiO line profiles.

Figure 16
shows the CS/SiO velocity versus the CO velocity for this sample of twelve
stars.  Overall, the CO lines are slightly broader than the CS/SiO
lines, probably due to radiative transfer effects in the more extended
CO envelope. Additionally, the 
stars fall into two groups; three carbon stars with $\rm V_o
(CO)$ = $\rm V_o(CS)$ and $\rm V_o ~ > ~ 10 ~ km~s^{-1}$; and the remaining
nine stars in which $\rm V_o(CS, ~ SiO) ~ < ~ V_o(CO)$ and $\rm V_o
~ < ~ 10 ~ km~s^{-1}$.  The second group contains both carbon and 
oxygen stars and stars with both parabolic and double line profiles.

CO is abundant and has effective self-shielding against photodestruction;
further, its rotational levels are relatively easily excited.  $\rm
^{29}SiO$ and CS have much lower abundances and also need higher
densities to produce significant rotational line emission; accordingly, 
the CO emission preferentially measures gas at much larger distances
from the star than do lines of less abundant molecules.  
In these stars, then, the dense gas in the
inner regions of the envelope, which gives rise to the CS and $\rm 
^{29}SiO$ lines, may not yet be accelerated to the terminal velocity.
The results in Figure 16 suggest
that the winds from these circumstellar envelopes do not reach their
terminal outflow velocity until distances of a few $\rm \times 10^{15}$ cm
from the star.
\vfil
\eject

\parindent=0pc
\line{\bf References \hfil}

\ref

Alcolea, J., Bujarrabal, V., \& S\'anchez Contreras, C. 1996,
  A\&A, 312, 560

\ref
Bachiller, R., Huggins, P.J., Cox, P., \& Forveille, T. 1991, A\&A, 247, 525

\ref
Bujarrabal, V., Fuente, A., \& Omont, A. 1994, ApJ, 421, L47

\ref
Cernicharo, J., Gu\'elin, M., Pe\~nalver, J., Martin-Pintado, J., 
  \& Mauersberger, R. 1989, A\&A, 222, L1

\ref
Claussen, M.J., Kleinmann, S.G., Joyce, R.R., \& Jura, M.
  1987, ApJS, 65, 385

\ref
Cohen, M., \& Hitchon, K. 1996, AJ, 111, 962

\ref
Crosas, M., \& Menten, K. 1997, ApJ, 483, 913

\ref
Dickinson, D.F., \& Chaisson, E.J. 1973, ApJ, 181, L135

\ref
Dickinson, D.F., \& Dinger, A.S.C. 1982, ApJ, 254, 136

\ref
Frogel, J.A., Persson, S.E., \& Cohen, J.G. 1980, ApJ, 239, 495

\ref
Gammie, C.F., Knapp, G.R., Young, K., Phillips, T.G., \& Falgarone, E.
  1989, ApJL, 345, L87

\ref
Gezari, D.Y., Schmitz, M., Pitts, P.S., \& Mead, J.M. 1993,``Catalogue of
  Infrared Observations'', NASA Reference Publication 1294

\ref
Gilman, R.S. 1972, ApJ, 178, 423

\ref
Goldreich, P.G., \& Scoville, N.Z. 1976, ApJ, 205, 144

\ref
Groenewegen, M. A. T. 1994, A\&A, 290, 531

\ref
Habing, H.J. 1996, A\&A Rev, 7, 97

\ref
Huggins, P.J., \& Healy, A.P. 1986, ApJ, 304, 418

\ref
IRAS Explanatory Supplement, Second Edition,  1988, NASA Publications

\ref
IRAS Point Source Catalogue, Second Edition, 1988, NASA Publications

\ref
Ivezic, Z., \& Elitzur, M. 1995, ApJ, 445, 415

\ref
Izumiura, H., Hashimoto, O., Kawara, K., Yamamura, I., \& Waters, L.B.F.M.
  1996, A\&A, 315, L221

\ref
Jaminet, P.A., Danchi, W.C., Sandell, G., \& Sutton, E.C. 1992, ApJ, 400, 535

\ref
Jones, T.J., Bryja, C.O., Gehrz, R.D., Harrison, T.E., Johnson, J.J., 
  Klebe, D.I., \& Lawrence, G.F. 1990, ApJS, 74, 785

\ref
Jorissen, A., \& Knapp, G.R. 1997, A\&AS (in press)

\ref
Jura, M. 1988, ApJS, 66, 33

\ref
Justtanont, K., Barlow, M.J., Skinner, C.J., Roche, P.F., Aitken, D.K., 
  \& Smith, C.H. 1996, A\&A, 309, 612

\ref
Kahane, C., \& Jura, M. 1996, A\&A, 310, 952

\ref
Kastner, J. H. 1992, ApJ, 401, 337

\ref
Kerschbaum, F., \& Hron, J. 1992, A\&A, 263, 97

\ref
Kerschbaum, F., \& Hron, J. 1994, A\&AS, 106, 397

\ref
Kerschbaum, F., Lazaro, C., \& Harbison, P. 1996, A\&AS, 118, 397

\ref
Kholopov, P.N., Samus, N.N., Frolov, M.S.,
  et al. 1985, `General Catalogue of Variable Stars', Moscow,
  Nauka Publishing House, Vols 1-4

\ref
Knapp, G.R., Jorissen, A., \& Young, K. 1997, A\&A, 326, 318

\ref
Knapp, G.R., \& Morris, M. 1985, ApJ, 292, 640

\ref
Knapp, G.R., Stanek, K.Z., Bowers, P.F., Young, K., \& Phillips, T.G.
  1997a, in preparation

\ref
Knapp, G.R., Woodhams, M.D., Gammie, C.F., Young, K., \& Phillips, T.G. 
  1997b, in preparation

\ref
Knapp, G.R., Young, K., \& Sahai, R. 1998, in preparation

\ref
Le Bertre, T. 1992, A\&AS, 94, 377

\ref
Le Bertre, T. 1993, A\&AS, 97, 729

\ref
L\`ebre, A., Mauron, N., Gillett, D., \& Barth\`es, D. 1996, A\&A, 310, 923

\ref
van Leeuwen, F., Feast, M.W., Whitelock, P.A., \& Yudin, B. 1997, 
  MNRAS, 287, 955

\ref
Lindqvist, M., Lucas, R., Olofsson, H., Omont, A., Eriksson, K., \&
  Gustafsson, B. 1996, A\&A, 305, L57

\ref
Loup, C., Forveille, T., Omont, A., \& Paul, J.F. 1993, A\&AS, 99, 291

\ref
Mamon, G.A., Glassgold, A.E., \& Huggins, P.J. 1988, ApJ, 328, 797

\ref
Margulis, M., van Blerkom, D.J., Snell, R.L., \& Kleinmann, S.G. 1990, 
  ApJ, 361, 673

\ref
Morris, M. 1980, ApJ, 236, 823

\ref
Morris, M., Guilloteau, S., Lucas, R., \& Omont, A. 1987, ApJ, 321, 888

\ref
Neri, R., Garcia-Burillo, S., Gu\'elin, M., Cernicharo, J., Guilloteau,
  S., \& Lucas, R. 1992, A\&A, 262, 544

\ref
Neri, R., Kahane, C., Lucas, R., Bujarrabal, V., \& Loup, C. 1997,
 A\&AS (in press)

\ref
Netzer, N., \& Elitzur, M. 1993, ApJ, 410, 701

\ref
Nyman, L.A., Booth, R.S., Carlstrom, U., et al. 1992, A\&AS, 93, 121

\ref
Olofsson, H. 1996, Ap. Space Sci., 245, 169

\ref
Olofsson, H. 1997a, in I.A.U. Symposium 177, `The Carbon Star Phenomenon',
  ed. R.F. Wing, D. Reidel Co., in press.

\ref
Olofsson, H. 1997b, in IAU Symposium 178, `Molecules in Astrophysics: Probes
  and Processes', D. Reidel Co. (in press).

\ref
Olofsson, H., Bergman, P., Eriksson, K., \& Gustafsson, B. 1996, A\&A, 311, 587

\ref
Perryman, M.A.C., Lindegren, L., Kovalevsky, J.,  et al.
 1997, A\&A, 323, L49

\ref
Planesas, P., Bachiller, R., Martin-Pintado, J., \& Bujarrabal, V.
  1990a, ApJ, 351, 263

\ref
Planesas, P., Kenney, J.D.P., \& Bachiller, R. 1990b, ApJL,
  364, L9

\ref
Sahai, R. 1992, A\&A, 253, L33

\ref
Sahai, R., \& Nyman, L-\AA{}. 1997, ApJ, 487, L155

\ref
Skinner, C.J., Dougherty, S.M., Meixner, M., Bode, M.F., Davis, R.J., 
  Drake, S.A., Arens, J.F., \& Jernigan, J.G. 1997, MNRAS, 288, 295

\ref
van der Veen, W.E.C.J., \& Habing, H.J. 1988, A\&A, 194, 125

\ref
van Winckel, H. 1997, A\&A, 319, 561

\ref
van Winckel, H., Waelkens, C., \& Waters, L.B.F.M. 1996, A\&A, 306, L37

\ref
Waters, L.B.F.M., Loup, C., Kester, D.J.M., Bontekoe, Tj. R., \& de Jong, T.
  1994, A\&A, 281, L1

\ref
Whitelock, P.A., Feast, M.W., Marang, F., \& Overbeek, M.D. 1997,
  MNRAS, 288, 512

\ref
Yamamura, I., Shibata, K.M., 
  Kasuga, T., \& Deguchi, S. 1994, ApJ,
  427, 406

\ref
Young, K. 1995, ApJ, 445, 872

\ref
Young, K., Phillips, T.G., \& Knapp, G.R. 1993a, ApJS, 86, 517

\ref
Young, K., Phillips, T.G., \& Knapp, G.R. 1993b, ApJ, 409, 725

\ref
Young, K., Serabyn, G., Phillips, T.G., Knapp, G.R., G\"usten, R.,
  \& Schulz, A. 1992, ApJ, 385, 265

\ref
Zuckerman, B., Dyck, H.M., \& Claussen, M.J. 1986, ApJ, 304, 401

\bigskip
\vfil
\eject
\centerline{Table 1. Observed Stars}
\bigskip
$$\vbox{
\tabskip 1em plus 2em
\halign to \hsize{\hfil$\rm #$\hfil& & \hfil$\rm #$\hfil \cr
Star& \alpha(1950)& \delta(1950)& Sp.& Var& Chem& P(days)& Profile\cr
& & & & & & & \cr
IRC+40004& 00~04~17.7& +42~48~18& M10& Mira& O& & P\cr
R~And& 00~21~23.0& +38~18~02.3& S4,6e& Mira& S& 409& (P)\cr
R~Scl& 01~24~40.02& -32~48~06.8& C6,4& SRb& C& 370& A/Blue\cr
IRC+50049& 01~55~37.3& +45~11~31.9& M8& & O& & D\cr
o~Cet& 02~16~49.04& -03~12~13.4& M7e& Mira& O& 332& D\cr  
IRC+50096& 03~22~59.1& +47~21~22.0&  C& Mira& C& 540& P\cr
IRC+60144& 04~30~45.9& +62~10~12& & & C& & (D)\cr
TX~Cam& 04~56~40.6& +56~06~28& M8.5& Mira& O& 557& P\cr
R~Lep& 04~57~19.7& -14~52~47.5& C7,4e& Mira& C& 433& P\cr
\alpha~Ori& 05~52~27.8& +07~23~57.9& M2Iab& SRc& O& 2070& A/Blue\cr
U~Ori& 05~52~51.0& +20~10~06.2& M8& Mira& O& 372& P\cr
AFGL~865& 06~01~17.4& +07~26~06.0& & Mira& C& 696& P\cr
AFGL~971& 06~34~16.5& +03~28~04.0& & Mira& C& 653& (P)\cr
GX~Mon& 06~50~03.5& +08~29~02& M9& Mira& O& 527& P\cr
R~Gem& 07~04~20.78& +22~46~56.7& S3,9e& Mira& S& 370& D\cr
HD~56126& 07~13~25.3& +10~05~09.2& F5Iab:& & C& 27& P\cr
S~CMi& 07~30~00.3& +08~25~35.5& M7e& Mira& O& 332& (P)\cr
RS~Cnc& 09~07~37.7& +31~10~05& M6se& SRc?& S& 120& D\cr
R~LMi& 09~42~34.7& +34~44~34.3& M7e& Mira& O& 372& A/Red\cr
IW~Hya& 09~42~56.4& -21~47~53.6& M9& Mira& O& 636& P\cr
R~Leo& 09~44~52.2& +11~39~41.9& M8e& Mira& O& 313& A/Red\cr
}}$$
\vfil
\eject
\centerline{Table 1, continued}
\bigskip
$$\vbox{
\tabskip 1em plus 2em
\halign to \hsize{\hfil$\rm #$\hfil& & \hfil$\rm #$\hfil \cr
Star& \alpha(1950)& \delta(1950)& Sp.& Var& Chem& P(days)& Profile\cr
& & & & & & & \cr
IRC+10216& 09~45~14.89& +13~30~40.8& C9,5& Mira& C& 637& P\cr
U~Hya& 10~35~05.0& -13~07~26.2& C6,4& SRb& C& 450& P\cr
R~Crt& 10~58~06.0& -18~03~21.8& M7& SRb& O& 160& P\cr
BK~Vir& 12~27~48.1& +04~41~34.5& M7& SRb& O& & P\cr
Y~UMa& 12~38~04.4& +56~07~14.8& M7& SRb& O& 168& P\cr
Y~CVn& 12~42~47.08& +45~42~47.9& C5,5J& SRb& C& 158& P\cr
RT~Vir& 13~00~05.7& +05~27~15.2& M8& SRb& O& 155& P\cr
SW~Vir& 13~11~29.75& -02~32~32.6& M7& SRb& O& 150& A/Red\cr
R~Hya& 13~26~58.48& -23~01~24.5& M6.5e& Mira& O& 388& D\cr
S~Vir& 13~30~23.2& -06~56~18.1& M7e& Mira& O& 378& (P)\cr
W~Hya& 13~46~12.2& -28~07~06.5& M8e& SRa& O& 382& P\cr
RX~Boo& 14~21~56.7& +25~55~48.5& M7.5& SRb& O& & P\cr
S~CrB& 15~19~21.53& +31~32~46.5& M7e& Mira& O& 360& P\cr
V~CrB& 15~47~44.08& +39~43~22.7& C6,2e& Mira& C& 358& (P)\cr
R~Ser& 15~48~23.2& +15~17~02.7& M6.5e& Mira& O& 357& (P)\cr
X~Her& 16~01~08.8& +47~22~35.8& M6& SRb& O& 95& D\cr
RU~Her& 16~08~08.6& +25~11~59& M7e& Mira& O& 485& P\cr
W~Aql& 19~12~41.68& -07~08~08.4& S3,9& Mira& S& 490& P\cr
\chi~Cyg& 19~48~38.48& +32~47~11.8& S7,2e& Mira& S& 407& P\cr
V~Cyg& 20~39~41.32& +47~57~45.4& C5,3e& Mira& C& 421& P\cr
EP~Aqr& 21~43~56.46& -02~26~40.9& M8& SRb& O& 55& D\cr
}}$$
\vfil
\eject
\centerline{Table 1, continued}
\bigskip
$$\vbox{
\tabskip 1em plus 2em
\halign to \hsize{\hfil$\rm #$\hfil& & \hfil$\rm #$\hfil \cr
Star& \alpha(1950)& \delta(1950)& Sp.& Var& Chem& P(days)& Profile\cr
& & & & & & & \cr
\Pi^1~Gru& 22~19~41.13& -46~12~02.4& S5,7& SRb& S& & D\cr
IRC+40540& 23~32~01.3& +43~16~27& C8,3.5& & C& 628& P\cr
R~Cas& 23~55~51.7& +51~06~36.4& M7e& Mira& O& 431& A/Red\cr
}}$$

\parindent 0pc
Notes to Table 1:

1. Spectral types, variable types, chemistry and periods mostly from Loup et al.
(1993) and from SIMBAD listings (periods primarily from Kholopov 
et al. 1985). Periods
for R Scl, R Lep, CRL 865, and CRL 971 from Le Bertre (1992).  
Period  for IRC+50096 from Jones et al. (1990). Period
for HD 56126 from L\`ebre et al. (1996).  Period for IW Hya from 
Le Bertre (1993).  Period for IRC+10216; mean of periods from 
Le Bertre (1992) and Cohen and Hitchon (1996).  Period of 
IRC+40540 from Cohen and Hitchon (1996).

2. Mira (o Cet): CO profile classification from 
data given by Planesas et al. (1993a,b) and Knapp et al. (1997).
$\Pi^1$ Gru: CO profile classification from data given by Sahai (1992).

3. CO line profiles for CRL 971, IRC+60144 and V Cyg affected by Galactic
emission. 

4. CO line profile classification:

P: parabolic.  Classifications in parentheses are for profiles with low
signal-to-noise ratio,  significant Galactic contamination or uncertain
classification.

D: double; two distinct components centered at the same velocity with
different outflow velocities.

A: asymmetric, with  Red or Blue emission wings.

\vfil
\eject
\centerline{Table 2. CO Line Parameters from High Velocity Resolution
Profiles}
\bigskip
$$\vbox{
\tabskip 1em plus 2em
\halign to \hsize{\hfil$\rm #$\hfil& & \hfil$\rm #$\hfil \cr
Star& line& r.m.s.& I_{CO}& T_{peak}& V_c& V_o\cr
& & & & & & \cr
IRC+40004& 2-1& 0.064& 14.0 \pm 2.6&  0.51 \pm 0.08& -20.5 \pm 1.8&
19.6 \pm 2.4\cr
R~And& 3-2& 0.46~& 12.8 \pm 3.6& 1.07 \pm 0.63& -15.3 \pm 5.4&
~8.8 \pm 4.5\cr
R~Scl& 2-1& 0.017& 27.8 \pm 0.5& 1.01 \pm 0.10& -19.0 \pm 0.9& 
16.5 \pm 1.1\cr
IRC+50049& 2-1& 0.055& 8.1 \pm 1.0& 0.50 \pm 0.10& -2.5 \pm 0.5&
9.5 \pm 0.5\cr
& & & & 0.52 \pm 0.08& -1.8 \pm 0.2& 2.6 \pm 0.3\cr
o~Cet& 3-2& 0.23& 76.2 \pm 3.0& 5.8 \pm 1.2& +46.0 \pm 1.0& 6.7 \pm 1.0\cr
& & & & 9.0 \pm 1.4& +46.6 \pm 0.2& 2.4 \pm 0.4\cr
IRC+50096& 3-2& 0.21& 32.8 \pm 6.3& 1.64 \pm 0.25& -16.4 \pm 1.6& 
14.9 \pm 2.7\cr
IRC+60144& 2-1& 0.050& (14.3)& 0.45 \pm 0.08& -48.0 \pm 2.0&
20.5 \pm 2.1\cr
& & & & 0.16 \pm 0.06& -45.0 \pm 1.5& 5.6 \pm 2.4\cr
TX~Cam& 3-2& 0.17& 32.7 \pm 6.5& 1.27 \pm 0.21& +11.2 \pm 2.1&
19.0 \pm 3.2\cr
R~Lep& 2-1& 0.046& 13.2 \pm 1.6& 0.44 \pm 0.06& +12.3 \pm 1.2& 17.4 
\pm 1.5\cr
\alpha ~ Ori& 2-1& 0.039& 8.2 \pm 1.1& 0.32 \pm 0.06& +3.4 \pm 1.2&  
14.2 \pm 1.2\cr
U~Ori& 3-2& 0.10&  7.2 \pm 1.5& 0.71 \pm 0.18& -38.1 \pm 1.3& 
7.5 \pm 1.2\cr
AFGL~865& 2-1& 0.041& 16.5 \pm 1.4& 0.81 \pm 0.06& +43.1 \pm 0.7& 
15.3 \pm 0.9\cr
AFGL~971& 3-2& 0.14& (15.4)&  0.77 \pm 0.19& +3.3 \pm 2.9&
14.9 \pm 5.1\cr
GX~Mon& 2-1& 0.052& 22.9 \pm 2.0& 0.75 \pm 0.06& -9.2 \pm 1.0& 
19.5 \pm 1.2\cr
R~Gem& 2-1& 0.038& 2.68 \pm 0.84& 0.05 \pm 0.02& -59.2 \pm 0.7&
11.0 \pm 1.1\cr
& & & & 0.30 \pm 0.05& -59.0 \pm 0.6& 4.8 \pm 0.9\cr
HD~56126& 2-1& 0.048& 9.2 \pm 0.9& 0.64 \pm 0.07& +73.0 \pm 0.8& 
10.7 \pm 1.1\cr
S~CMi& 3-2& 0.095& 1.5 \pm 0.6& 0.31 \pm 0.08& +51.5 \pm 1.4& 
3.3 \pm 1.4\cr
}}$$
\vfil
\eject
\centerline{Table 2, continued}
\bigskip
$$\vbox{
\tabskip 1em plus 2em
\halign to \hsize{\hfil$\rm #$\hfil& & \hfil$\rm #$\hfil \cr
Star& line& r.m.s.& I_{CO}& T_{peak}& V_c& V_o\cr
& & & & & & \cr
RS~Cnc& 2-1& 0.074& 12.5 \pm 1.2& 0.88 \pm 0.0.14& +7.5 \pm 0.5&
8.0 \pm 0.5\cr
& & & & 1.00 \pm 0.13& +7.2 \pm 0.2& 2.4 \pm 0.3\cr
& 3-2& 0.12& 18.0 \pm 1.9& 1.10\pm 0.2& +7.5 \pm 0.3&
8.0 \pm 0.6\cr
& & & & 1.66 \pm 0.2& +7.2 \pm 0.3& 2.8 \pm 0.4\cr
R~LMi& 2-1& 0.025& 2.72 \pm 0.38& 0.24 \pm 0.06& 0.0 \pm 0.6&
7.8 \pm 0.5\cr
IW~Hya& 3-2& 0.122& 12.3 \pm 3.4& 0.61\pm 0.18& +40.4 \pm 2.5&
13.6 \pm 4.1\cr
R~Leo& 3-2& 0.076& 21.9 \pm 1.2& 2.0 \pm 0.2& -0.5 \pm 0.5&
6.8 \pm 0.7\cr
IRC+10216& 2-1& 0.28& 424.2 \pm 8.3& 19.1 \pm 0.6& -25.5 \pm 0.3&
14.6 \pm 0.3\cr
U~Hya& 2-1& 0.044& 10.0 \pm 0.6& 0.98 \pm 0.10& -30.9 \pm 0.4&
6.7 \pm 0.5\cr
R~Crt& 2-1& 0.050& 11.1 \pm 1.1& 0.59 \pm 0.07& +11.2 \pm 0.7& 10.8
\pm 0.7\cr
BK~Vir& 2-1& 0.039& 8.4 \pm 0.5& 0.29 \pm 0.05& +17.1 \pm 0.8&
5.8 \pm 1.1\cr
Y~UMa& 2-1& 0.030& 3.2 \pm 0.3& 0.40 \pm 0.05& 19.0 \pm 0.5&
5.4 \pm 0.6\cr
Y~CVn& 3-2& 0.12& 10.7 \pm 1.9& 0.88 \pm 0.17& 21.1 \pm 0.9& 
7.8 \pm 1.3\cr
RT~Vir& 2-1& 0.023& 4.9 \pm 0.4& 0.37 \pm 0.03& +17.5 \pm 0.5&
8.9 \pm 0.6\cr
SW~Vir& 2-1& 0.022& 11.7 \pm 0.4& 0.89 \pm 0.11& -11.1 \pm 0.6& 
7.8 \pm 0.7\cr
R~Hya& 2-1& 0.022& 4.9 \pm 0.5& & & \cr
& 3-2& 0.10& 22.2 \pm 2.2& 1.2 \pm 0.3& -10.0 \pm 0.8& 
11.0 \pm 1.2\cr
& & & & 0.8 \pm 0.2& -9.7 \pm 0.6& 5.0 \pm 0.7\cr
S~Vir& 2-1& 0.025& 0.64 \pm 0.25& 0.09 \pm 0.03& +9.8 \pm 2.3&
5.2 \pm 2.2\cr
W~Hya& 3-2& 0.17& 28.3 \pm 2.7&  2.0 \pm 0.3& +40.8 \pm 0.6&
8.1 \pm 0.8\cr
}}$$
\vfil
\eject
\centerline{Table 2, continued}
\bigskip
$$\vbox{
\tabskip 1em plus 2em
\halign to \hsize{\hfil$\rm #$\hfil& & \hfil$\rm #$\hfil \cr
Star& line& r.m.s.& I_{CO}& T_{peak}& V_c& V_o\cr
& & & & & & \cr
RX~Boo& 3-2& 0.15& 23.9 \pm 2.4& 1.65 \pm 0.21&  +1.5 \pm 0.7&
9.6 \pm 1.0\cr
S~CrB& 2-1& 0.034& 2.53 \pm 0.51& 0.25 \pm 0.04& +1.1 \pm 0.9&
7.4 \pm 1.2\cr
V~CrB& 3-2& 0.23& 4.3 \pm 1.0& 0.39 \pm 0.08& -99.4 \pm 1.3&
6.3 \pm 1.2\cr
R~Ser& 3-2& 0.17& 3.5 \pm 1.9& 0.52 \pm 0.04& +31.6 \pm 0.5&
5.3 \pm 0.9\cr
X~Her& 2-1& 0.026& 6.4 \pm 0.4& 0.38 \pm 0.05& -72.8 \pm 0.8& 
8.5 \pm 1.0\cr
& & & & 0.45 \pm 0.07& -73.2 \pm 0.4& 3.2 \pm 0.5\cr
& 3-2& 0.071& 13.3 \pm 1.3& 0.7 \pm 0.1& -73.2 \pm 0.5& 
9.0 \pm 1.0\cr
& & & & 0.99 \pm 0.12& -73.1 \pm 0.3& 3.5 \pm 1.4\cr
RU~Her& 2-1& 0.011& 2.3 \pm 0.2& 0.18 \pm 0.03& -12.1 \pm 0.8& 
9.4 \pm 1.5\cr
W~Aql& 2-1& 0.030& 43.9 \pm 1.2& 1.63 \pm 0.09& -24.1 \pm 0.6&
17.6 \pm 0.8\cr
\chi~Cyg& 2-1& 0.038& 28.8 \pm 0.7& 1.98 \pm 0.14& +9.9 \pm 0.4&
8.9 \pm 0.5\cr
V~Cyg& 2-1& 0.067& (25.0)& 1.61 \pm 0.09& +14.0 \pm 0.5& 
11.8 \pm 0.6\cr
EP~Aqr& 2-1& 0.035& 12.8 \pm 0.8& 0.70 \pm 0.10& -33.0 \pm 0.5&
10.8 \pm 0.8\cr
& & & & 1.36 \pm 0.21& -33.8 \pm 0.2& 1.4 \pm 0.2\cr
& 3-2& 0.10& 24.3 \pm 2.3& 1.30 \pm 0.3& -33.0 \pm 0.5&
10.8 \pm 1.0\cr
& & & & 3.05 \pm 0.60& -33.7 \pm 0.2& 1.5 \pm 0.3\cr
\Pi^1~Gru& 2-1& -& -& -& -10& 27\cr
& & & & -& -13& 12\cr
IRC+40540& 3-2& 0.17& 49.8 \pm 4.8& 2.61 \pm 0.26& -16.8 \pm 0.9&
14.3 \pm 1.4\cr
R~Cas& 3-2& 0.20& 51.8 \pm 4.8& 2.94 \pm 0.34& +24.9 \pm 0.9& 
12.1 \pm 1.2\cr
}}$$
\vfil
\eject
\parindent 0pc
Notes:

1. IRC+60144, AFGL 971 and V Cyg:  contaminated by galactic emission

2. R Hya CO(2-1) - no fit possible

3. $\Pi^1$ Gru: data from Sahai (1992) observed with  15 m SEST
\vfil
\eject
\centerline{Table 3. Mass Loss Rates}
\bigskip
$$\vbox{
\tabskip 1em plus 2em
\halign to \hsize{\hfil$\rm #$\hfil& & \hfil$\rm #$\hfil \cr
Star& D(pc)& Chem& V_o (km~s^{-1})& \mathaccent 95 M (M_{\odot}~yr^{-1})& 
Dist. ~ Method\cr
& & & & & \cr
IRC+40004& 850& O& 19.6& 8.4 \times 10^{-6}& M_K, ~L_{bol}\cr
R~And& 490& S& ~8.8& 9.4 \times 10^{-7}& M_K\cr
R~Scl& 475& C& 16.5& 2.7 \times 10^{-6}& M_K, ~Hipparcos\cr
IRC+50049& 142& O& ~9.5& 1.5 \times 10^{-7}& Hipparcos\cr
& & & ~2.6& 2.7 \times 10^{-8}& \cr
o~Cet& 128& O& ~6.7& 4.4 \times 10^{-7}& Hipparcos\cr  
& & & ~2.4& 9.4 \times 10^{-8}& \cr
IRC+50096& 650& C& 14.9& 4.4 \times 10^{-6}& M_K, ~ L_{bol}\cr
IRC+60144& 850& C& 20.5& 4.8 \times 10^{-6}& M_K, ~ L_{bol}\cr
& & & 5.6& 2.4 \times 10^{-7}& \cr
TX~Cam& 280& O& 19.0& 2.3 \times 10^{-6}& M_K, L_{bol}\cr
R~Lep& 250& C& 17.4& 5.2 \times 10^{-7}& Hipparcos\cr
\alpha~Ori& 131& O& 14.2& 3.1 \times 10^{-7}& Hipparcos\cr
U~Ori& 300& O& ~7.5& 2.9 \times 10^{-7}& M_K, P-M_K, Hipparcos\cr
AFGL~865& 1800& C& 15.3& 2.2 \times 10^{-5}& M_K, ~ L_{bol}\cr
AFGL~971& 1300& C& 14.9& 7.6 \times 10^{-6}& M_K, L_{bol}\cr
GX~Mon& 500& O& 19.5& 4.8 \times 10^{-6}& M_K, L_{bol}\cr
R~Gem& 850& S& 11.0& 4.1 \times 10^{-7}& M_K\cr
& & & ~4.8& 4.5 \times 10^{-7}& \cr
S~CMi& 470& O& ~3.3& 6.6 \times 10^{-8}& P - M_K; M_K\cr
}}$$
\vfil
\eject
\centerline{Table 3, continued}
\bigskip
$$\vbox{
\tabskip 1em plus 2em
\halign to \hsize{\hfil$\rm #$\hfil& & \hfil$\rm #$\hfil \cr
Star& D(pc)& Chem& V_o(km~s^{-1})& \mathaccent 95 M (M_{\odot}~yr^{-1})& 
Dist. ~ Method\cr
& & & & & \cr
RS~Cnc& 122& S& ~8.0& 1.0 \times 10^{-7}&  Hipparcos\cr
& & & ~2.6& 2.3 \times 10^{-8}& \cr
R~LMi& 270& O& ~7.8& 1.6 \times 10^{-7}& P-M_K; M_K; Hipparcos\cr
IW~Hya& 900& O& 13.6& 4.2 \times 10^{-6}& M_K; L_{bol}\cr
R~Leo& 101& O& ~6.8& 9.4 \times 10^{-8}& Hipparcos\cr
IRC+10216& 150& C& 14.6& 9.2 \times 10^{-6}& M_K, L_{bol}\cr
U~Hya& 162& C& ~6.7& 9.2 \times 10^{-8}& Hipparcos\cr
R~Crt& 300& O& 10.8& 6.9 \times 10^{-7}& M_K, P-M_K, Hipparcos\cr
BK~Vir& 176& O& ~5.8& 6.9 \times 10^{-8}& Hipparcos\cr
Y~UMa& 313& O& ~5.4& 1.7 \times 10^{-7}& Hipparcos\cr
Y~CVn& 218& C& ~7.8& 1.1 \times 10^{-7}& Hipparcos\cr
RT~Vir& 138& O& ~8.9& 1.1 \times 10^{-7}& Hipparcos\cr
SW~Vir& 143& O& ~7.8& 1.7 \times 10^{-7}& Hipparcos\cr
R~Hya& 200& O& 11.0& 4.7 \times 10^{-7}& M_K; Hipparcos\cr
& & & ~5.0& 7.3 \times 10^{-8}& \cr
S~Vir& 420& O& ~5.2& 8.4 \times 10^{-8}& Hipparcos\cr
W~Hya& 115& O& ~8.1& 1.7 \times 10^{-7}& Hipparcos\cr
RX~Boo& 156& O& ~9.6& 3.0 \times 10^{-7}& Hipparcos\cr
S~CrB& 400& O& ~7.4& 2.5 \times 10^{-7}& M_K; P-M_K; Hipparcos\cr
V~CrB& 600& C& ~6.3& 2.5 \times 10^{-7}& M_K; Hipparcos\cr
}}$$
\vfil
\eject
\centerline{Table 3, continued}
\bigskip
$$\vbox{
\tabskip 1em plus 2em
\halign to \hsize{\hfil$\rm #$\hfil& & \hfil$\rm #$\hfil \cr
Star& D(pc)& Chem& V_o(km~s^{-1})& \mathaccent 95 M (M_{\odot}~yr^{-1})& 
Dist. ~ Method\cr
& & & & & \cr
R~Ser& 279& O& ~5.3&  9.2 \times 10^{-8}& Hipparcos\cr
X~Her& 138& O& ~9.0& 1.1 \times 10^{-7}& Hipparcos\cr
& & & ~3.4& 3.4 \times 10^{-8}& \cr
RU~Her& 400& O& ~9.4& 3.2 \times 10^{-7}& M_K; P-M_K; Hipparcos\cr
W~Aql& 610& S& 17.6& 9.4 \times 10^{-6}& M_K\cr
\chi~Cyg& 106& S& ~9.0& 2.3 \times 10^{-7}& Hipparcos\cr
V~Cyg& 271&  C& 11.8& 9.4 \times 10^{-7}& Hipparcos\cr
EP~Aqr& 135& O& 10.8& 2.3 \times 10^{-7}& Hipparcos\cr
& & & ~1.4& 1.7 \times 10^{-8}& \cr
\Pi^1~Gru& 153& S& 27~& 5.7 \times 10^{-7}& Hipparcos\cr
& & & 12~& 4.6 \times 10^{-7}& \cr
IRC+40540& 700& C& 14.3& 7.4 \times 10^{-6}& M_K; L_{bol}\cr
R~Cas& 107& O& 12.1& 5.2 \times 10^{-7}& Hipparcos\cr
}}$$

\vfil
\eject
\centerline{Table 4. Photodissociation Radii of Stars with Double Winds}
\bigskip
$$\vbox{
\tabskip 1em plus 2em
\halign to \hsize{\hfil$\rm #$\hfil& & \hfil$\rm #$\hfil \cr
Star& V_o(km~s^{-1})& R_p(cm)& T(years)\cr
& & & \cr
o~Cet& ~6.7& 3.8\times 10^{16}& 1800\cr
& ~2.4& 2.4 \times 10^{16}& 3100\cr
IRC+60144& 20.5& 1.4\times 10^{17}& 2170\cr
& ~5.6& 1.7 \times 10^{17}& 9700\cr
R~Gem& 11.0& 3.5 \times 10^{16}& 1000\cr
& ~4.8& 5.2 \times 10^{16}& 3390\cr
RS~Cnc& ~8.0& 1.8 \times 10^{16}& 710\cr
& ~2.6& 1.1 \times 10^{16}& 1370\cr
R~Hya& 11.0& 3.3 \times 10^{16}& 950\cr
& ~5.0& 6.0 \times 10^{16}& 3750\cr
X~Her& ~9.0& 1.7 \times 10^{16}& 580\cr
& ~3.4& 1.1 \times 10^{16}& 1050\cr
EP~Aqr& 10.8& 2.3 \times 10^{16}& 670\cr
& ~1.4& 1.1 \times 10^{16}& 2390\cr
\Pi^1~Gru& 27& 3.6 \times 10^{16}& 420\cr
& 12& 3.6 \times 10^{16}& 950\cr
}}$$
\vfil
\eject
\centerline{Table 5. Search for Fast Molecular Winds in Observed Stars}
\bigskip
$$\vbox{
\tabskip 1em plus 2em
\halign to \hsize{\hfil$\rm #$\hfil& & \hfil$\rm #$\hfil \cr
Star& Line& T(fast)& R\cr
& & & \cr
IRC+40004& 2-1& <0.015& <0.028\cr
RAnd& 3-2& <0.12& <0.12\cr
IRC+50049& 2-1& <0.01& <0.02\cr
IRC+50096& 3-2& <0.04& <0.09\cr
IRC+60144& 2-1& <0.015& <0.03\cr
TX~Cam& 3-2& <0.05& <0.04\cr
R~Lep& 2-1& (0.02)& (0.05)\cr
\alpha~Ori& 2-1& <0.01& <0.03\cr
U~Ori& 3-2& <0.03& <0.04\cr
AFGL~865& 2-1& <0.015& <0.02\cr
AFGL~971& 3-2& <0.04& <0.05\cr
GX~Mon& 2-1& <0.015& <0.02 \cr
R~Gem& 2-1& <0.01& <0.05\cr
HD~56126& 2-1& <0.01& <0.02\cr
S~CMi& 3-2& <0.03& <0.2\cr
RS~Cnc& 2-1& <0.03& <0.02\cr
RS~Cnc& 3-2& <0.03& <0.02\cr
R~LMi& 2-1& <0.01& <0.04\cr
IW~Hya& 3-2& <0.025& <0.04\cr
R~Leo& 3-2& <0.025& <0.01\cr
IRC+10216& 2-1& (0.1)& (0.0048)\cr
}}$$
\vfil
\eject
\centerline{Table 5, continued}
\bigskip
$$\vbox{
\tabskip 1em plus 2em
\halign to \hsize{\hfil$\rm #$\hfil& & \hfil$\rm #$\hfil \cr
Star& Line& T(fast)& R\cr
& & & \cr
U~Hya& 2-1& <0.01& <0.01\cr
R~Crt& 2-1& <0.01& <0.02\cr
BK~Vir& 2-1& <0.01& <0.06\cr
Y~UMa& 2-1& <0.015& <0.06\cr
Y~CVn& 3-2& <0.03& <0.04\cr
RT~Vir&  2-1& <0.02& <0.062\cr
SW~Vir& 2-1& <0.02& <0.026\cr
R~Hya& 2-1& <0.02& <0.06\cr
R~Hya& 3-2& <0.035& <0.02\cr
S~Vir& 2-1& <0.01& <0.13\cr
W~Hya& 3-2& <0.03& <0.02\cr
RX~Boo& 3-2& (0.05)& (0.03)\cr
S~CrB& 2-1& <0.02& <0.10\cr
V~CrB& 3-2& <0.06& <0.14\cr
R~Ser& 3-2& <0.03& <0.07\cr
X~Her& 2-1& <0.03& <0.06\cr
X~Her& 3-2& <0.03& <0.02\cr
RU~Her& 2-1& <0.01&  <0.06\cr
W~Aql& 2-1& <0.03&<0.017\cr
\chi~Cyg& 2-1& <0.02& <0.009\cr
V~Cyg& 2-1& <0.06& <0.031\cr
}}$$
\vfil
\eject
\centerline{Table 5, continued}
\bigskip
$$\vbox{
\tabskip 1em plus 2em
\halign to \hsize{\hfil$\rm #$\hfil& & \hfil$\rm #$\hfil \cr
Star& Line& T(fast)& R\cr
& & & \cr
EP~Aqr& 2-1& <0.02& <0.022\cr
IRC+40540& 3-2& <0.03& <0.01\cr
R~Cas& 3-2& <0.04& <0.01\cr
AOS~comb& -& -& 0.0061\cr
}}$$
\bigskip
\bigskip
\parindent=0pc
Note: (1) T(fast) = brightness temperature of fast wind.
R = ratio of brightness temperatures of fast
and steady wind. Doubtful values in parentheses.

(2) Wings on the strong emission line from IRC+10216 are likely
to be instrumental - compare the data for the AOS frequency calibration
comb. 
\vfil
\eject
\centerline{Table 6. Wind Velocities in the CO(3-2), CS(7-6)
and $\rm ^{29}SiO(8-7)$ Lines}
\bigskip
\line{a. Carbon Stars\hfil}
\bigskip
$$\vbox{
\tabskip 1em plus 2em
\halign to \hsize{\hfil$\rm #$\hfil& & \hfil$\rm #$\hfil \cr
Star& V_o (CO 3-2)& V_o (CS 7-6)& Profile& V_o (CO 3-2)\cr
& & & & (50 ~MHz)\cr
& & & & \cr
IRC+50096& 15.2 \pm 0.7& 13.7 \pm 1.5& P& 14.9\pm 2.7\cr
AFGL 971& 15.3 \pm 0.8& 14.4 \pm 4.4& (P)& 14.9 \pm 5.1\cr
Y~CVn& 8.7 \pm 0.6& 7.2 \pm 2.2& P& 7.8 \pm 1.3\cr
V~CrB& 8.0 \pm 1.5& 3.9 \pm 0.9& (P)& 6.3 \pm 1.2\cr
IRC+40540& 14.7 \pm 0.5& 15.7 \pm 1.6& P& 14.3 \pm 1.4\cr
}}$$
\vfil
\eject
\centerline{Table 6, continued}
\line{b. Oxygen Stars \hfil}
$$\vbox{
\tabskip 1em plus 2em
\halign to \hsize{\hfil$\rm #$\hfil& & \hfil$\rm #$\hfil \cr
Star& V_o (CO 3-2)& V_o (^{29}SiO8-7)& Profile& V_o (CO 3-2)\cr
& & & & (50 ~MHz)\cr
& & & & \cr
U~Ori& 7.8 \pm 1.0& 5.6 \pm 1.7& P& 7.5 \pm 1.2\cr
RS~Cnc& 6.4 \pm 1.7& 4.2 \pm 0.2& D& 8.0 \pm 0.6\cr
& & & & 2.8 \pm 0.4\cr
R~Leo& 7.7 \pm 0.1& 4.7 \pm 0.7& A/red& 6.8 \pm 0.7\cr
R~Hya& 8.9 \pm 1.5& 3.7 \pm 1.7& D& 11.0 \pm 1.2\cr
& & & & 5.0 \pm 0.7\cr
RX~Boo& 10.0 \pm 0.5& 7.1 \pm 1.1& P& 9.6 \pm 1.0\cr
X~Her& 7.3 \pm 1.9& 5.7 \pm 1.1& D& 9.0 \pm 1.0\cr
& & & & 3.5 \pm 1.4\cr
R~Cas& 12.6 \pm 0.6& 5.2 \pm 1.0& A/red& 12.1 \pm 1.2\cr
}}$$
\vfil
\eject

\parindent=0pc

\line{\bf Figure Captions\hfil}

\underbar{Figure 1:} CO line profiles observed at a velocity resolution of
$\rm \sim ~ 0.1 ~ - ~ 0.2 ~ km~s^{-1}$ for 43 evolved stars.

\underbar{Figure 2:} Simulated CO(2-1) line profiles as observed with the
CSO 10.4 m telescope for stars with double winds.  In all cases the
faster wind has an outflow speed $\rm V_f ~ = ~ 10 ~ km~s^{-1}$ and 
a mass loss rate of $\rm 3 \times 10^{-7} ~ M_{\odot}~yr^{-1}$.
The model star is at a distance of 200 pc. The simulated line profiles
have a velocity resolution of 0.1 $\rm km~s^{-1}$ and an r.m.s.
channel-to-channel noise of 0.04 K.  The line profile of the fast
wind is parabolic with a peak temperature of 0.65 K.  (a) outflow
speed of slow wind $\rm V_s ~ = ~ 3 ~ km~s^{-1}$, peak temperature
$\rm T_s ~ = ~ 0.2 K$ (five times the r.m.s. noise).  (b) $\rm
V_s ~ = ~ 6 ~ km~s^{-1}$, $\rm T_s$ = 0.6 K.  (c) $\rm V_s ~ = ~
8 ~ km~s^{-1}$, $\rm T_s$ = 1.0 K.

\underbar{Figure 3:} Period distribution for the evolved stars in
Table 1.  Shaded columns: stars with double molecular winds.

\underbar{Figure 4:} [25] - [60] vs. [12] - [25] and K - [60] vs.
[12] - [25] color-color diagrams for evolved stars. Figures 3a and 3c:
colors as a function of line shape. 
Figures 3b and 3d: colors as a function of chemistry.  

\underbar{Figure 5:} CO profiles for stars with double winds showing
profile fits (dashed lines).

\underbar{Figure 6:} Histograms of wind outflow speeds for (a) the
broad component of stars with double winds (b) the narrow components of
stars with double winds and (c) stars with single winds.  The summed 
distributions for the broad and narrow components is shown by
dashed lines.

\underbar{Figure 7:}  Outflow speeds of slow versus fast winds for 9 stars
with double winds.

\underbar{Figure 8:} Absolute K magnitudes for stars with Hipparcos parallaxes
(a) observed (b) corrected for circumstellar and interstellar extinction.

\underbar{Figure 9:} Distributions of mass loss rates for (a) broad
components and (b) narrow components of stars with double winds: and (c) 
stars with single winds.

\underbar{Figure 10:} Mass loss rate versus wind outflow speed.  Filled
symbols: stars with Hipparcos parallaxes. Open symbols: distances derived
by other methods (see text).

\underbar{Figure 11:} Mass loss rate versus wind outflow speed.  Filled
symbols: stars with double winds, broad components.  Open symbols: 
stars with double winds, narrow components.  Crosses: stars with single
winds.

\underbar{Figure 12:} Mass loss rate versus wind outflow speed.
Filled symbols: carbon stars. Open symbols: oxygen stars.  Crosses:
S stars.

\underbar{Figure 13:} Mass loss rate in $\rm M_{\odot} ~ yr^{-1}$
versus pulsation period (days) for stars with measured periods (Table
1), sorted by variable type.  Filled symbols: Mira variables.  Open
symbols: SRb variables. Crosses: SRa, SRc and unknown type.  The
data for $\alpha$ Ori (SRb: P = 2070 days) are not plotted.  Vertical
lines join the mass loss rates for stars with double winds.

\underbar{Figure 14:} Wind outflow speed $\rm V_o$ in $\rm km~s^{-1}$
versus pulsation period (days) for stars with measured periods (Table
1), sorted by variable type.  Filled symbols: Mira variables.  Open
symbols: SRb variables. Crosses: SRa, SRc and unknown type.  The
data for $\alpha$ Ori (SRb: P = 2070 days) are not plotted.  Vertical
lines join the velocities for stars with double winds.

\underbar{Figure 15:}  CO line profiles, binned to a resolution of $\rm
\sim 7 ~ km~s^{-1}$, for (a) R Lep and (b) RX Boo.

\underbar{Figure 16:} Wind outflow speed measured from the CO(3-2)
line versus that measured from the $\rm ^{29}SiO(8-7)$ line (oxygen stars,
open symbols) or the CS(7-6) line (carbon stars, filled symbols).

\vfil
\eject
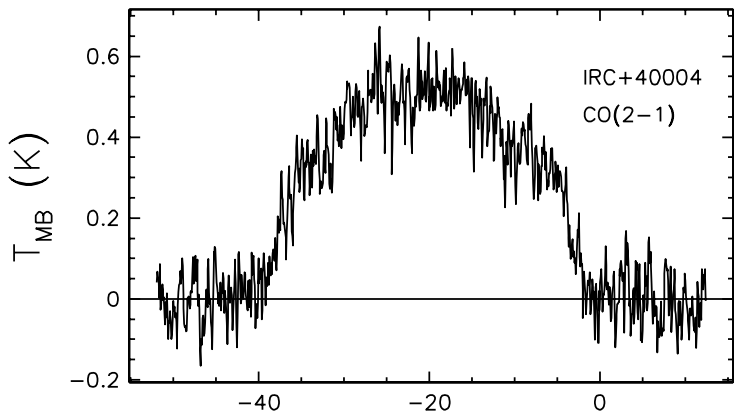
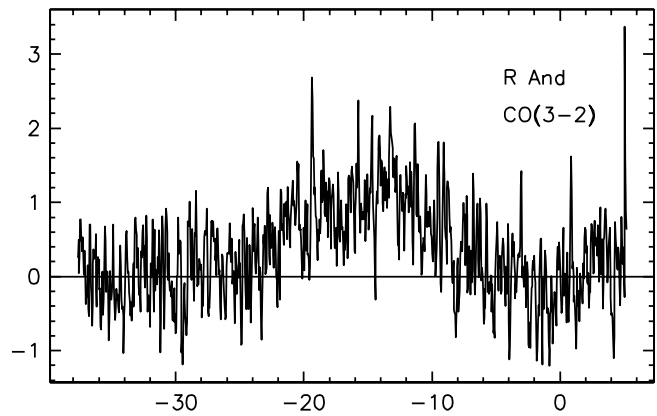
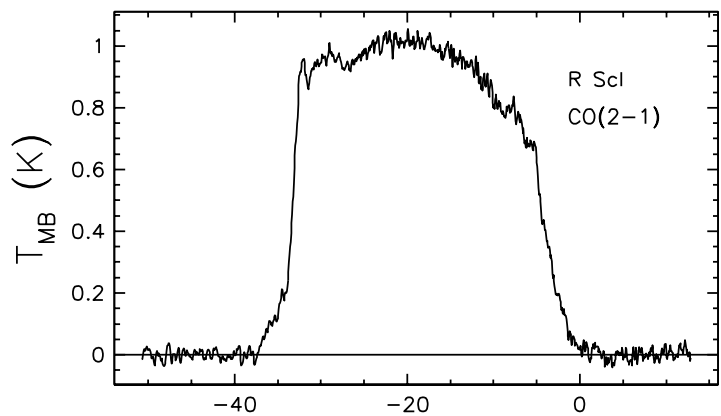
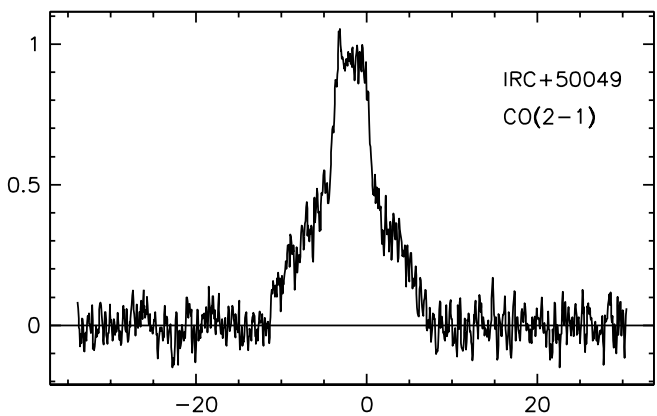
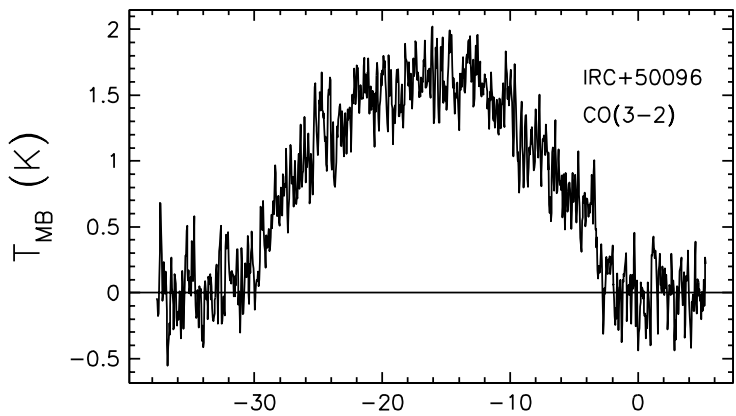
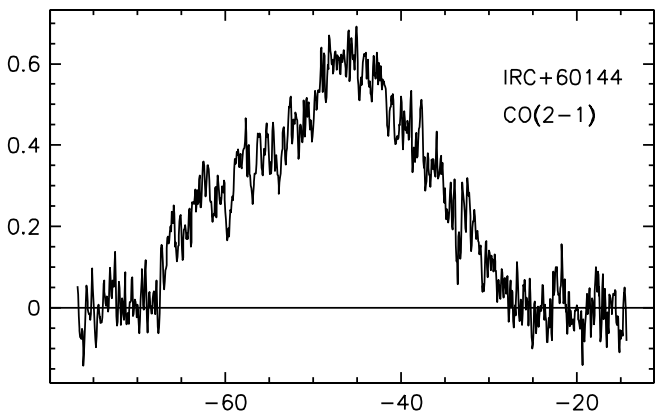
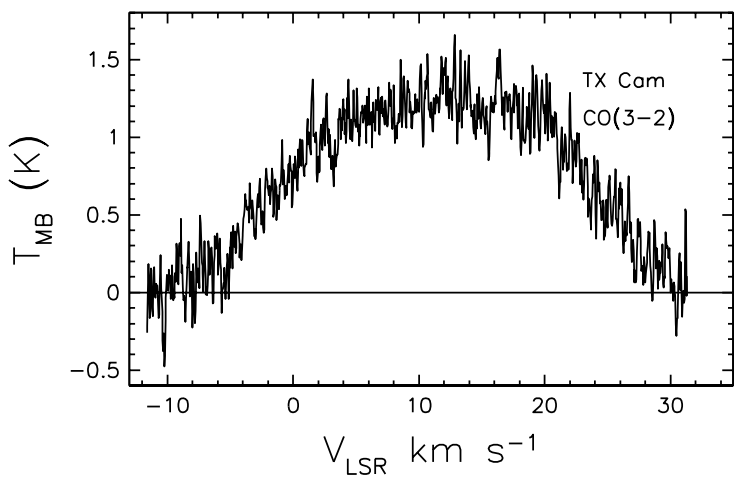
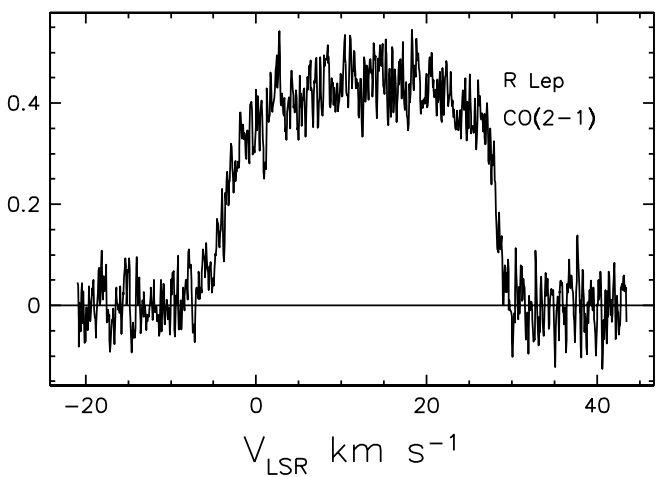
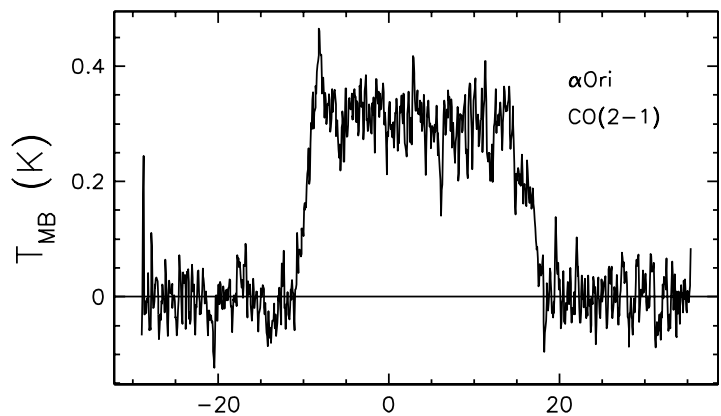
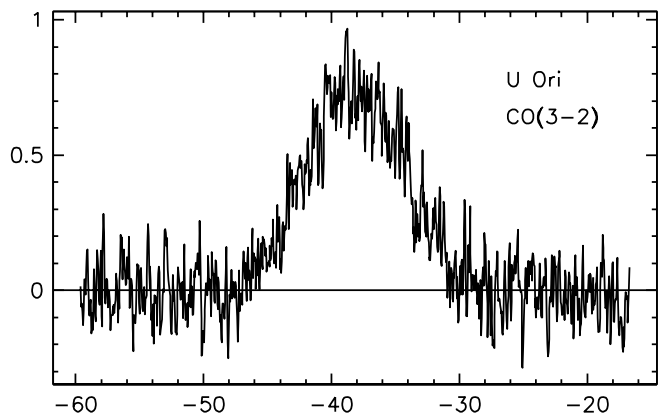
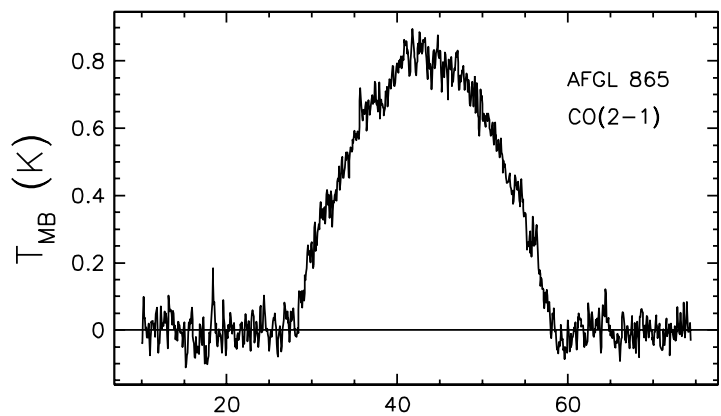
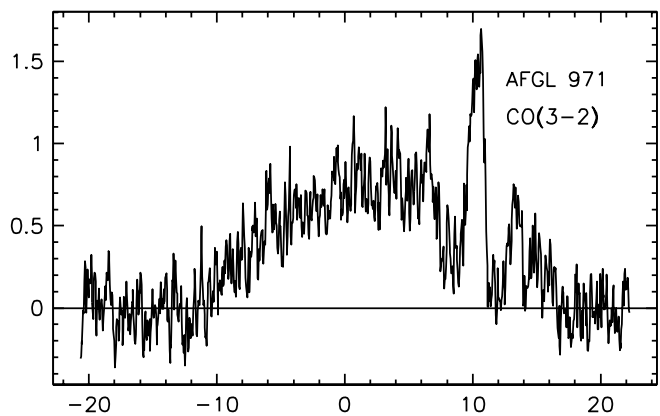
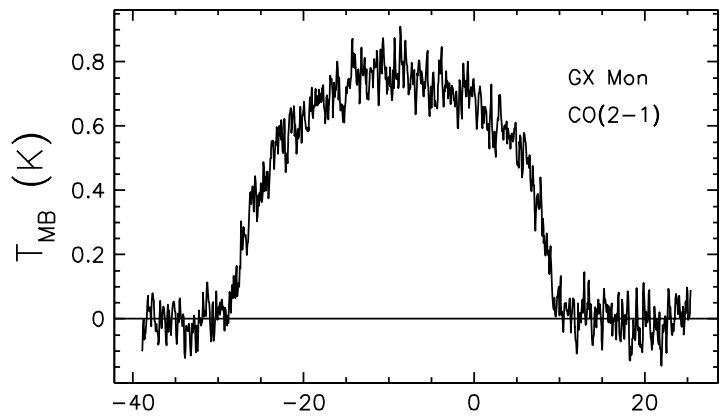
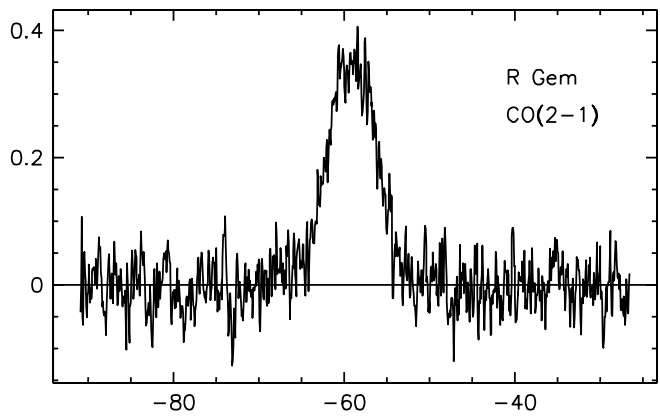
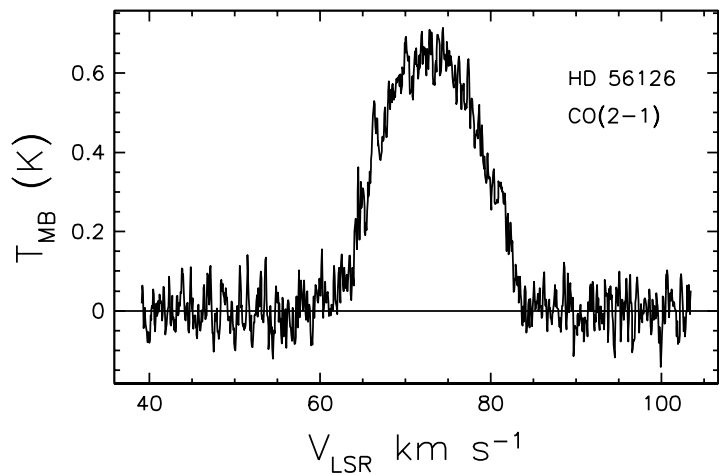
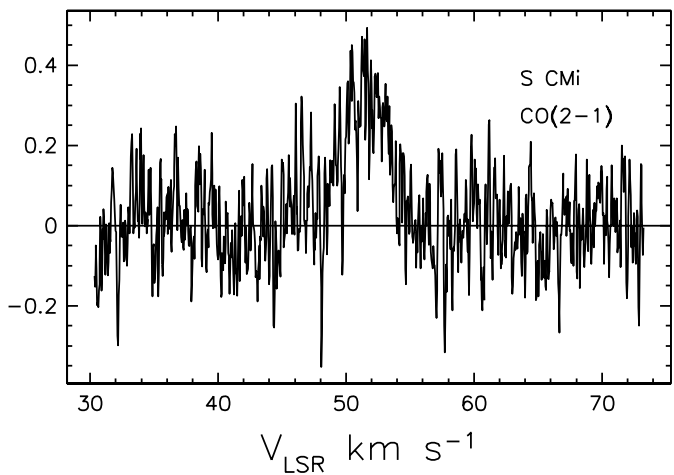
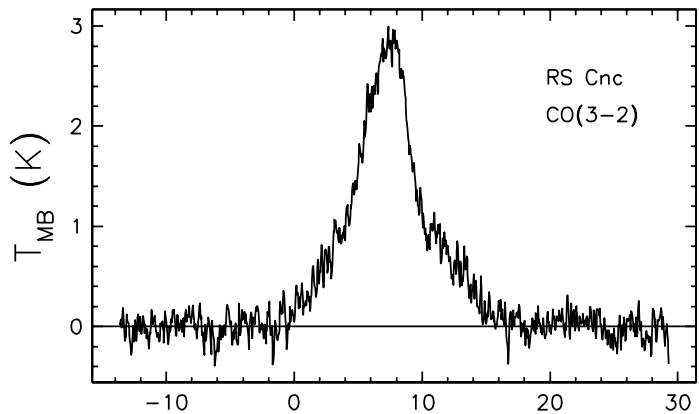
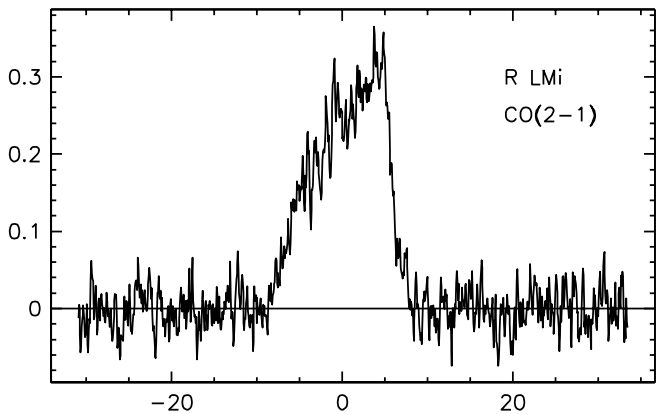
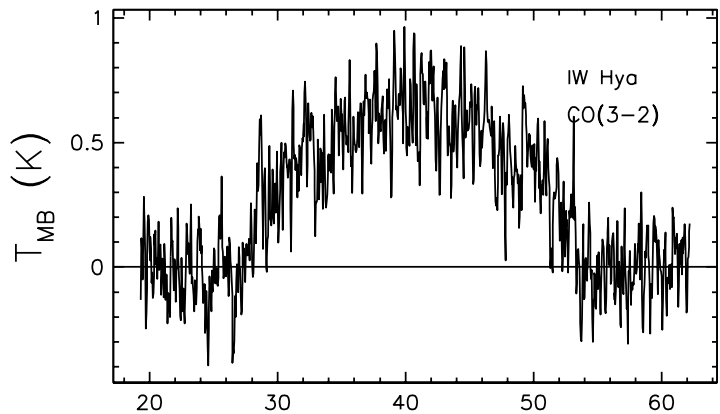
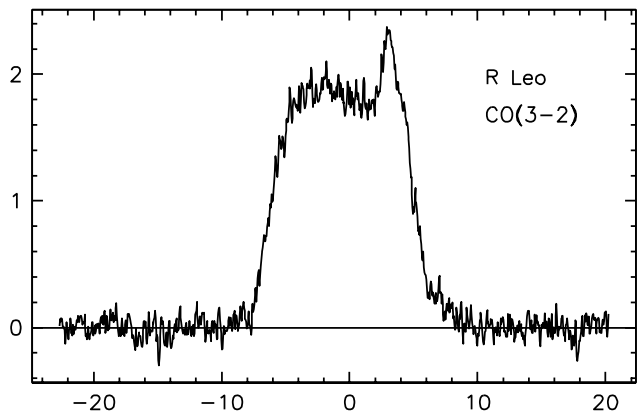
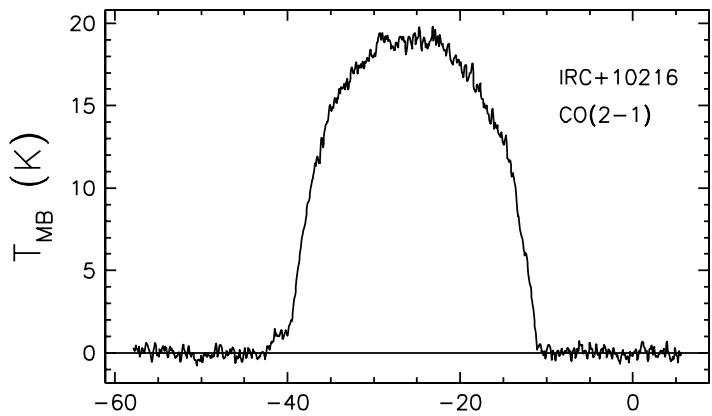
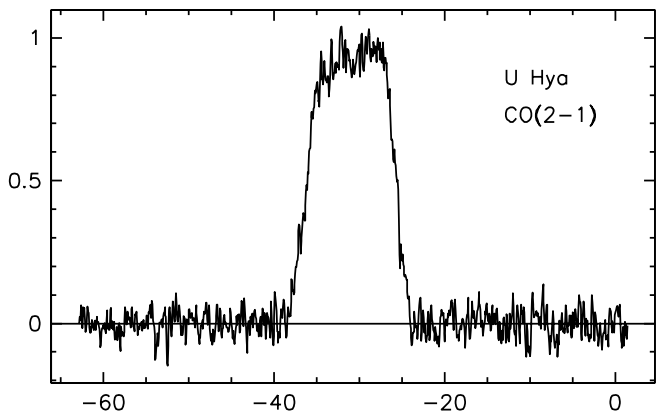
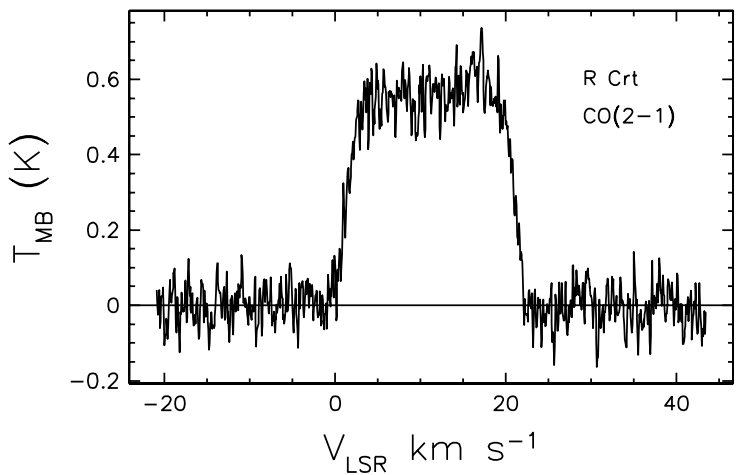
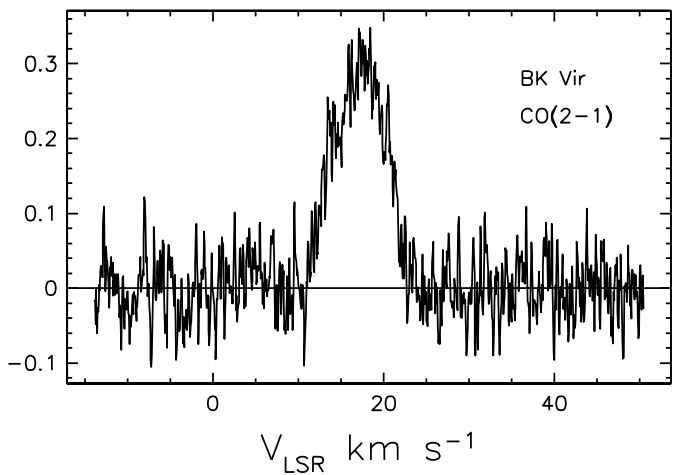
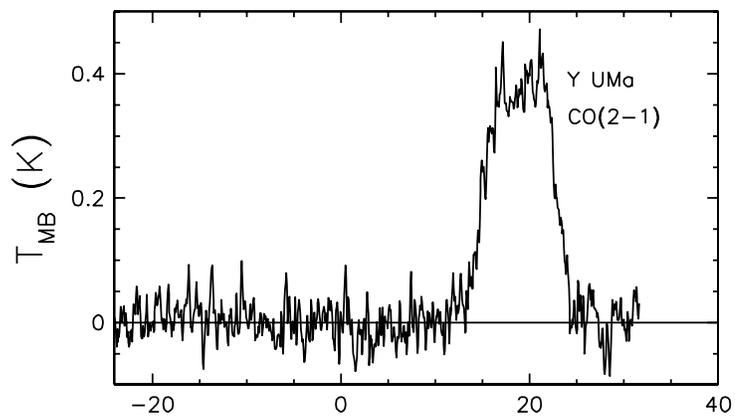
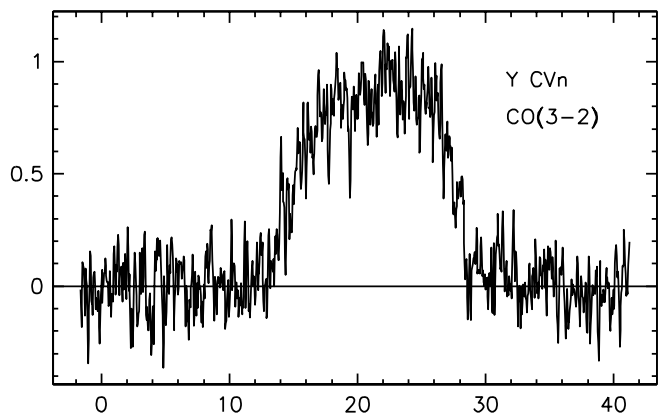
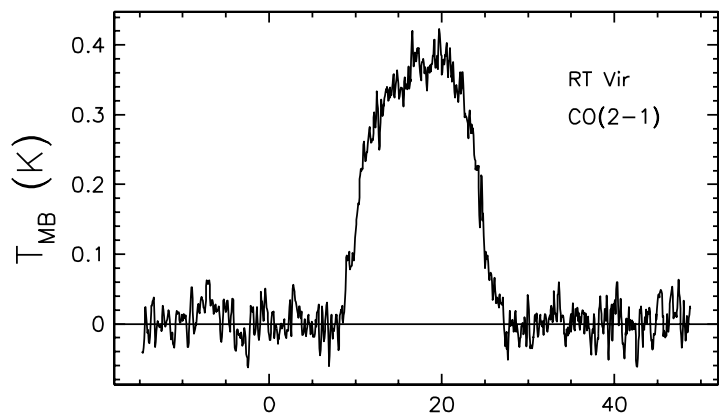
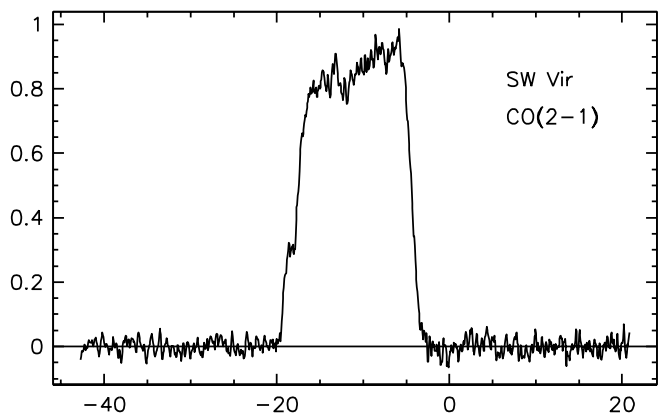
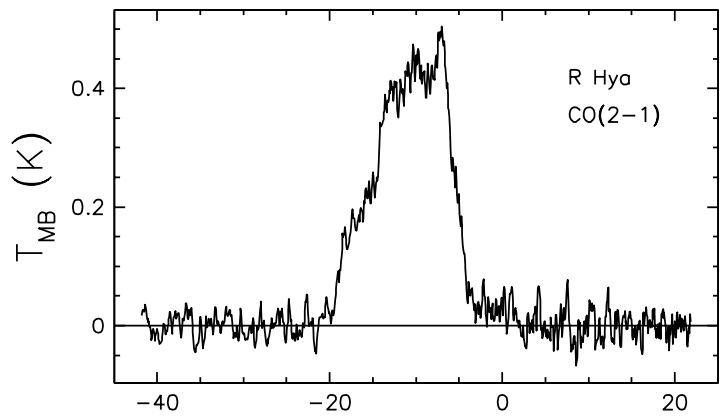
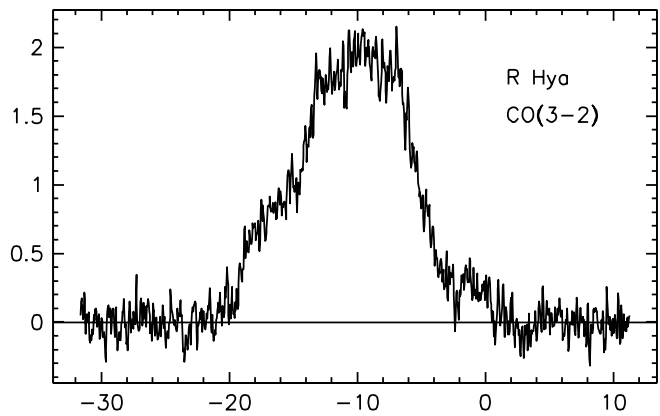
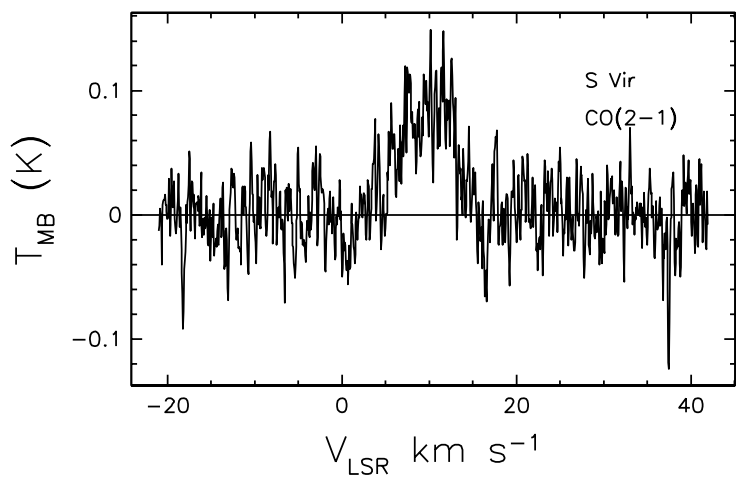
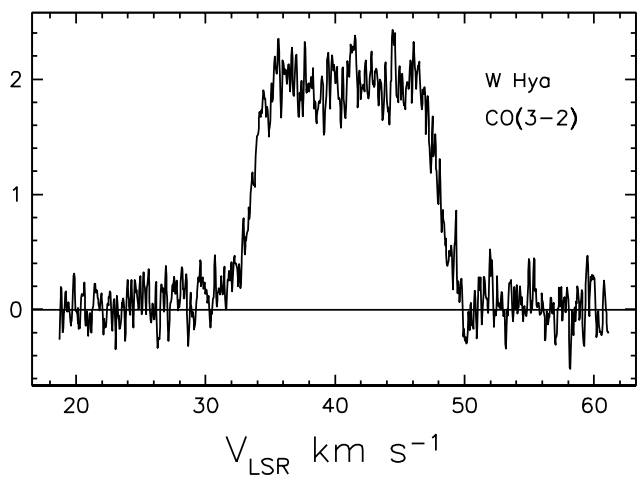
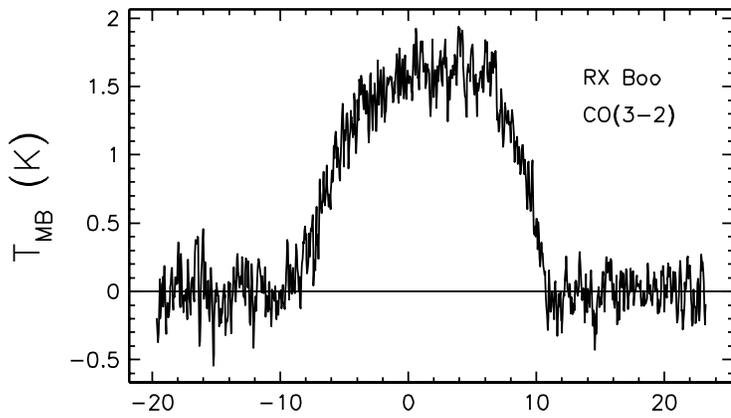
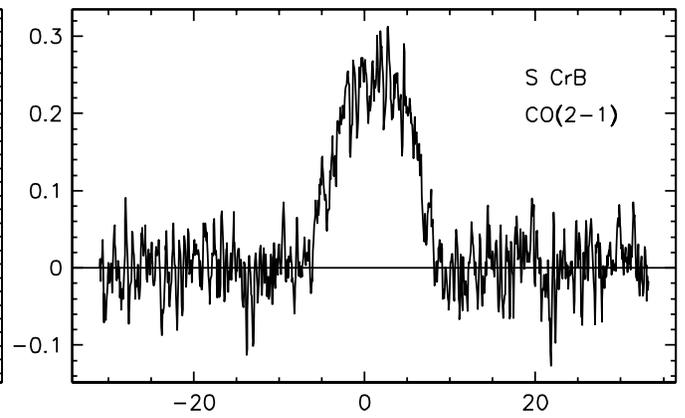
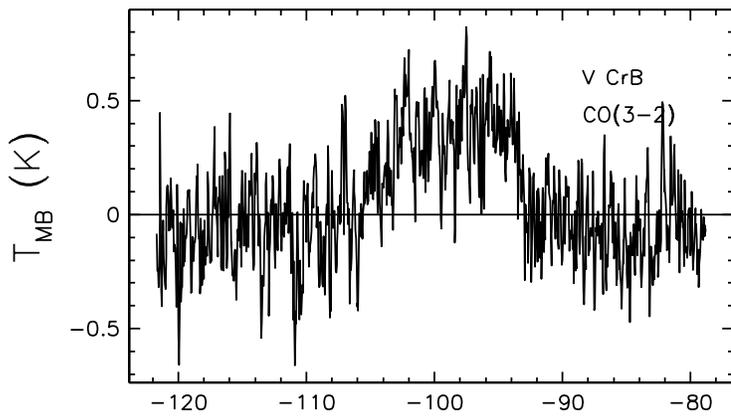
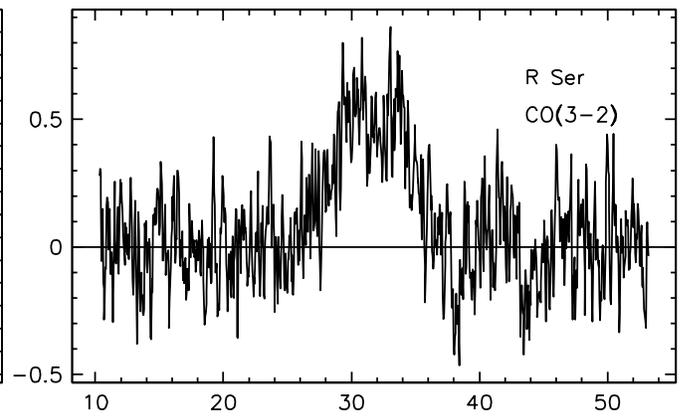
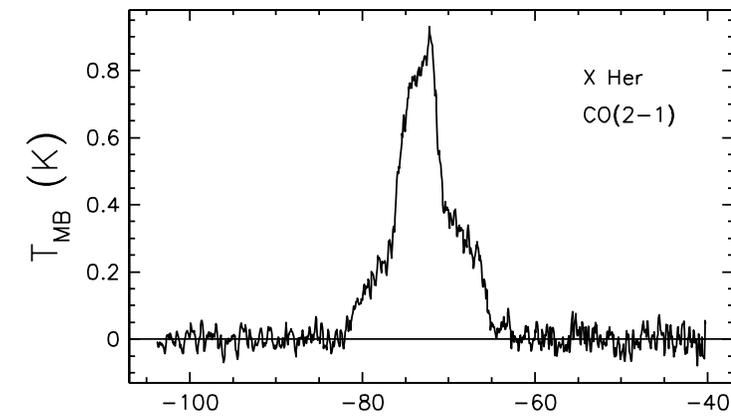
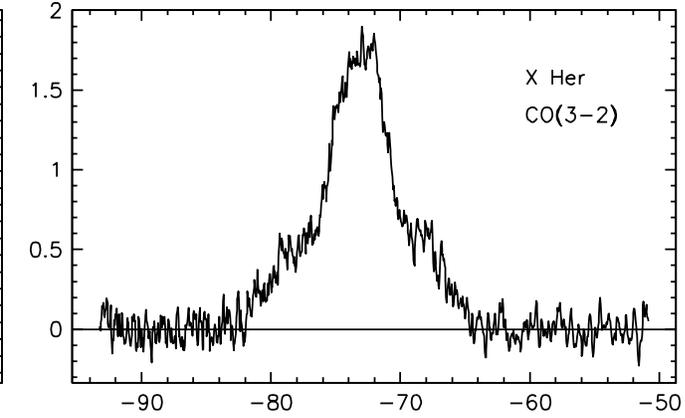
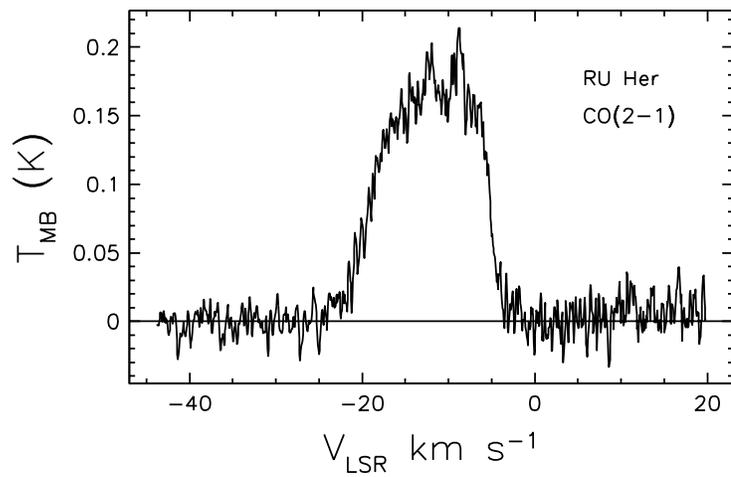
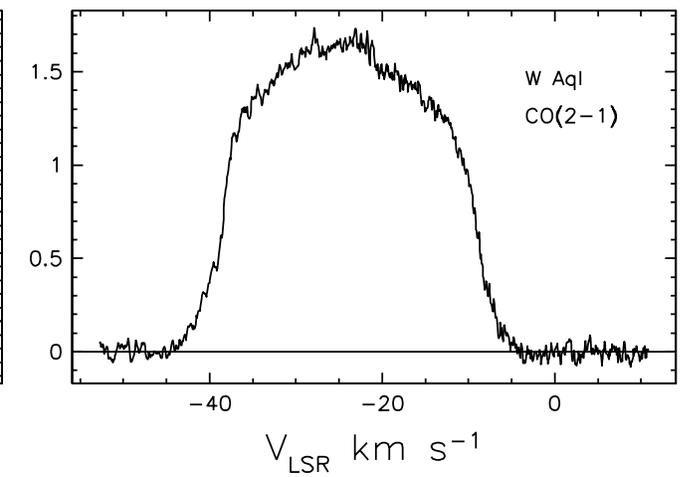
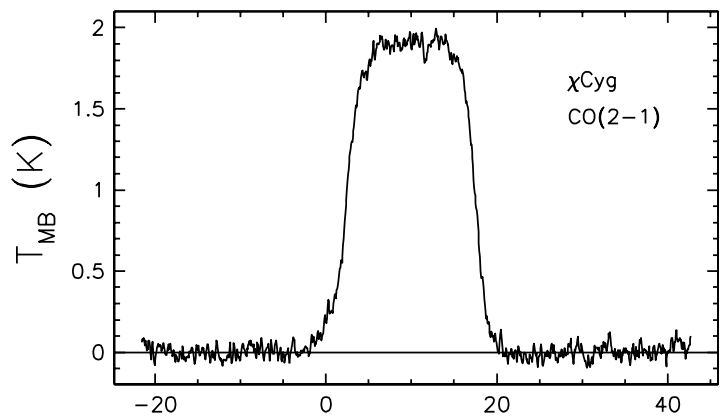
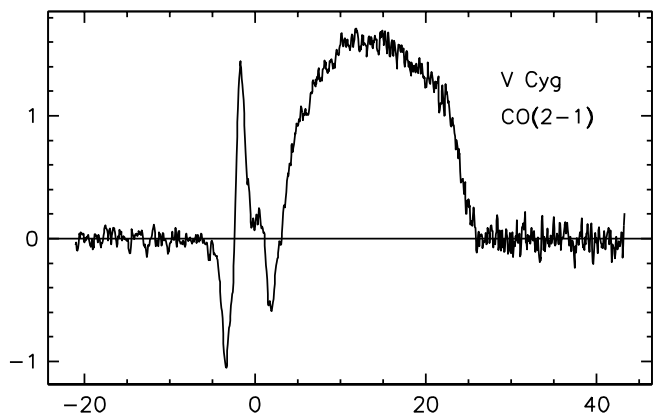
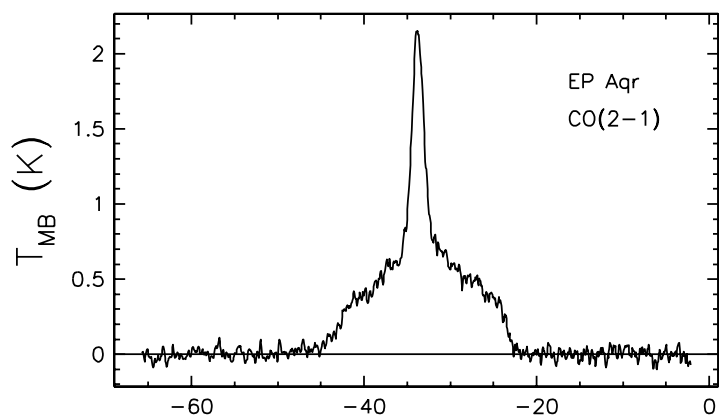
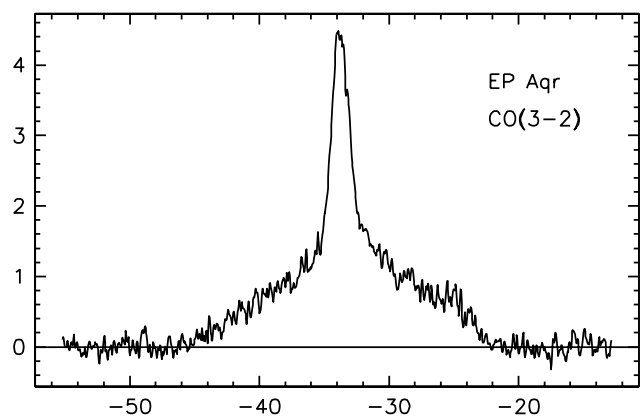
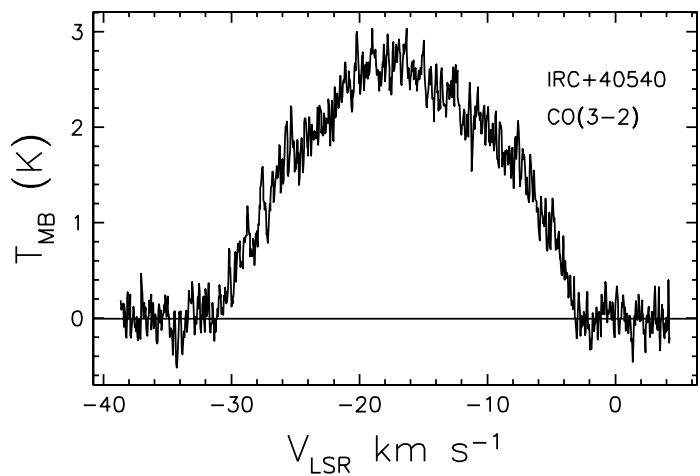
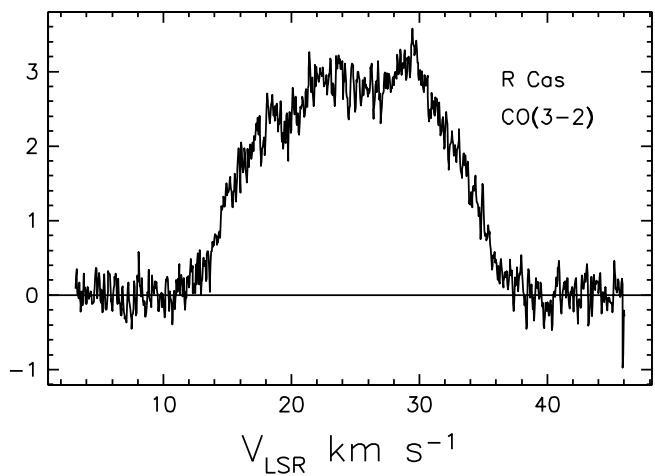
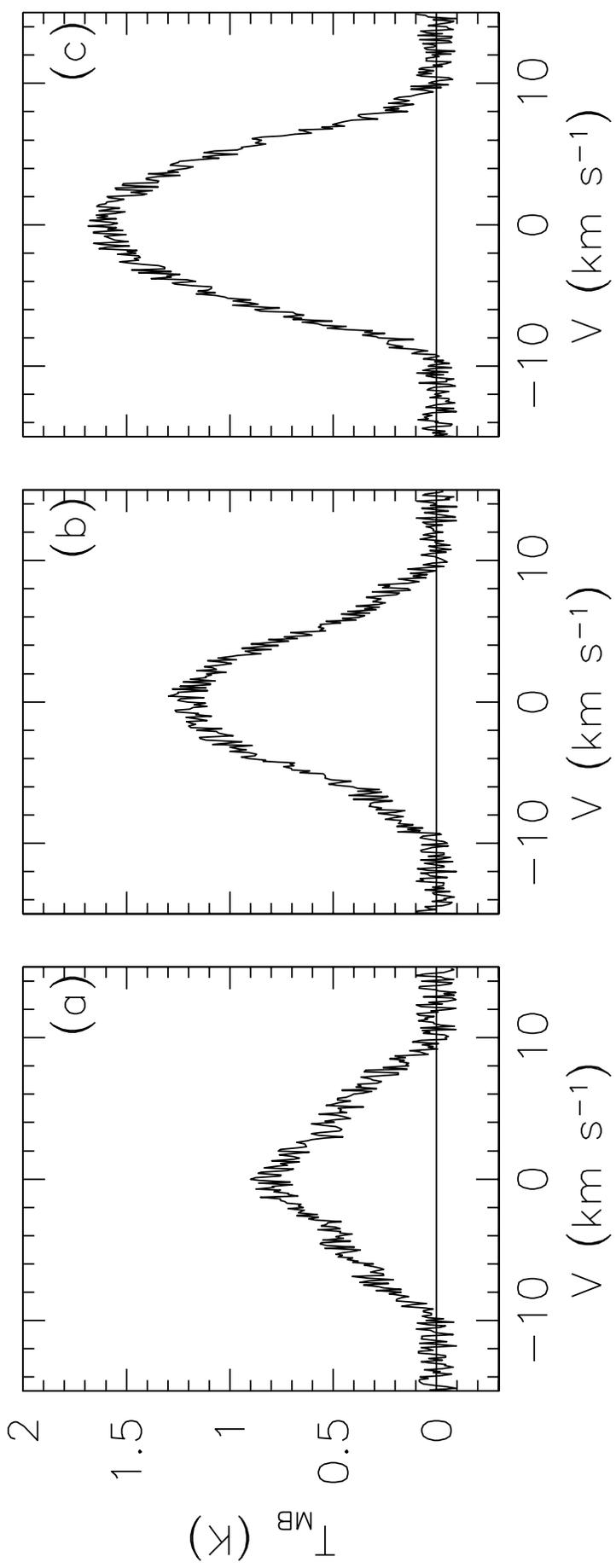
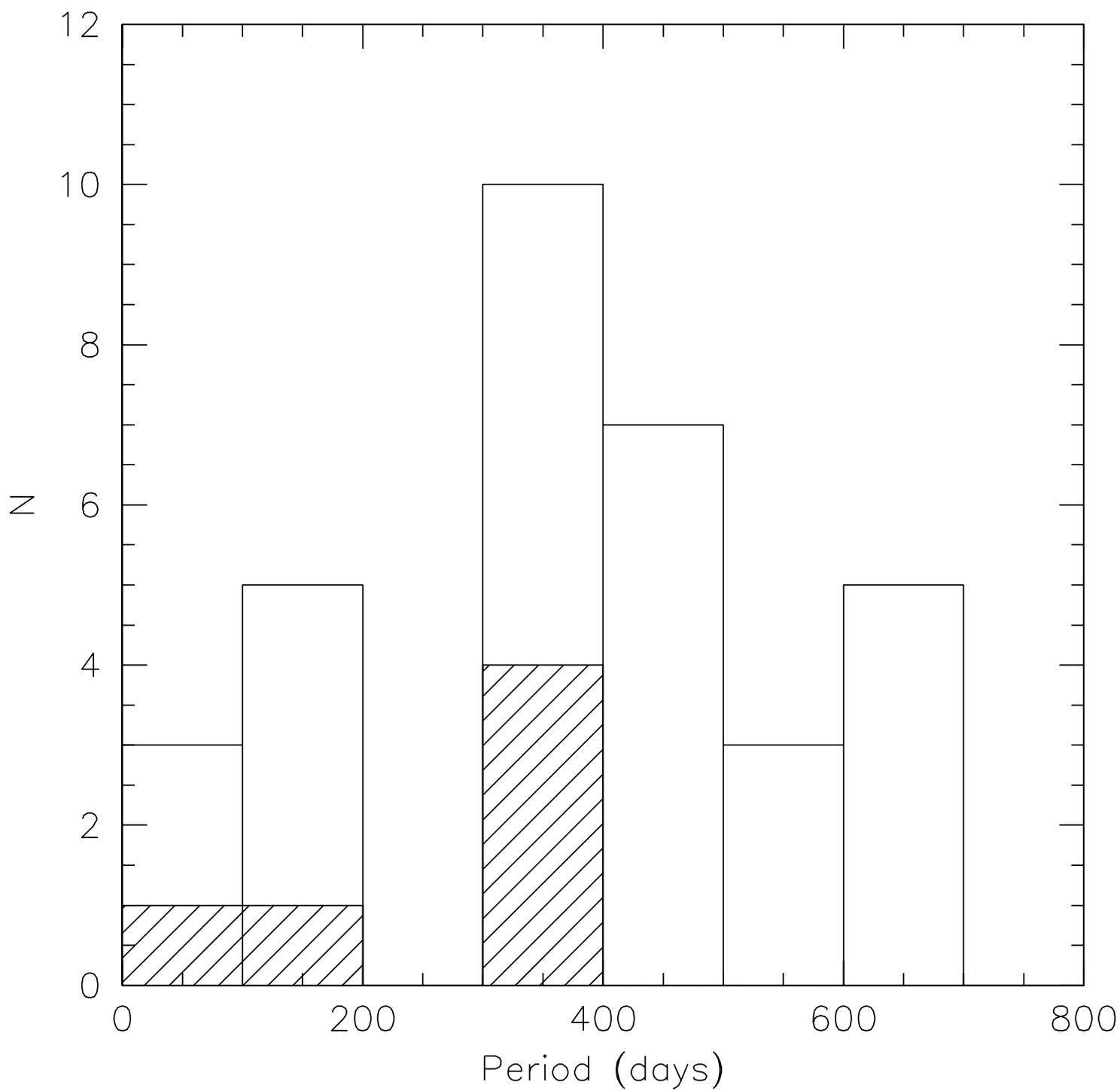
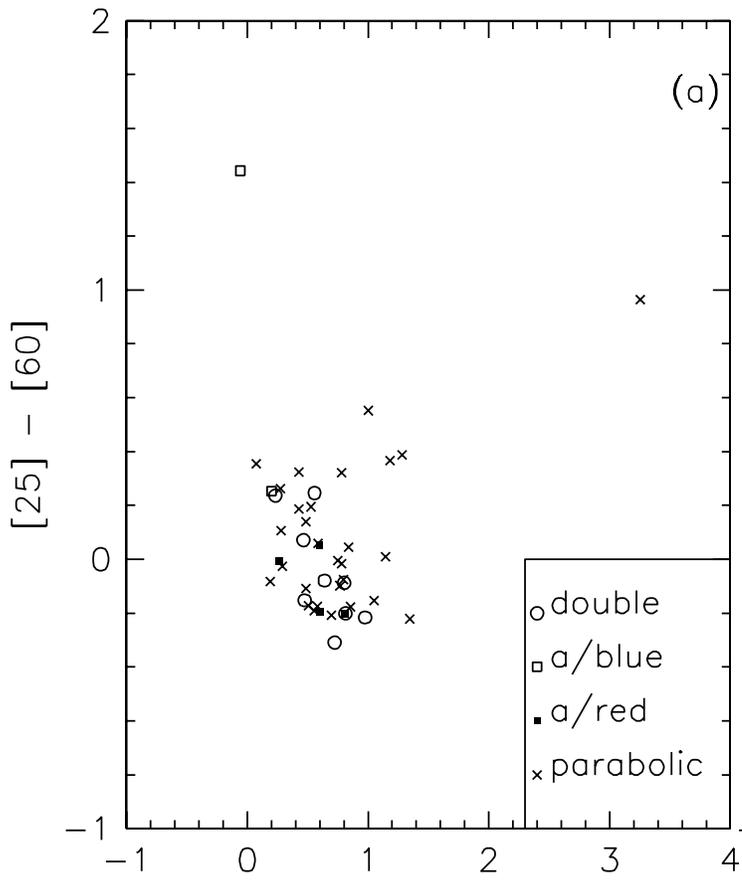
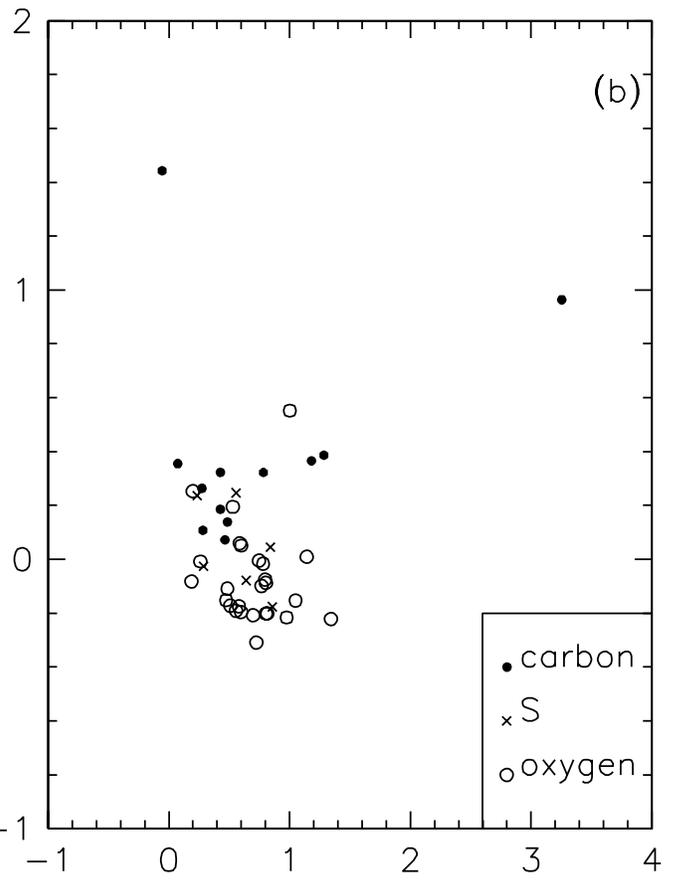
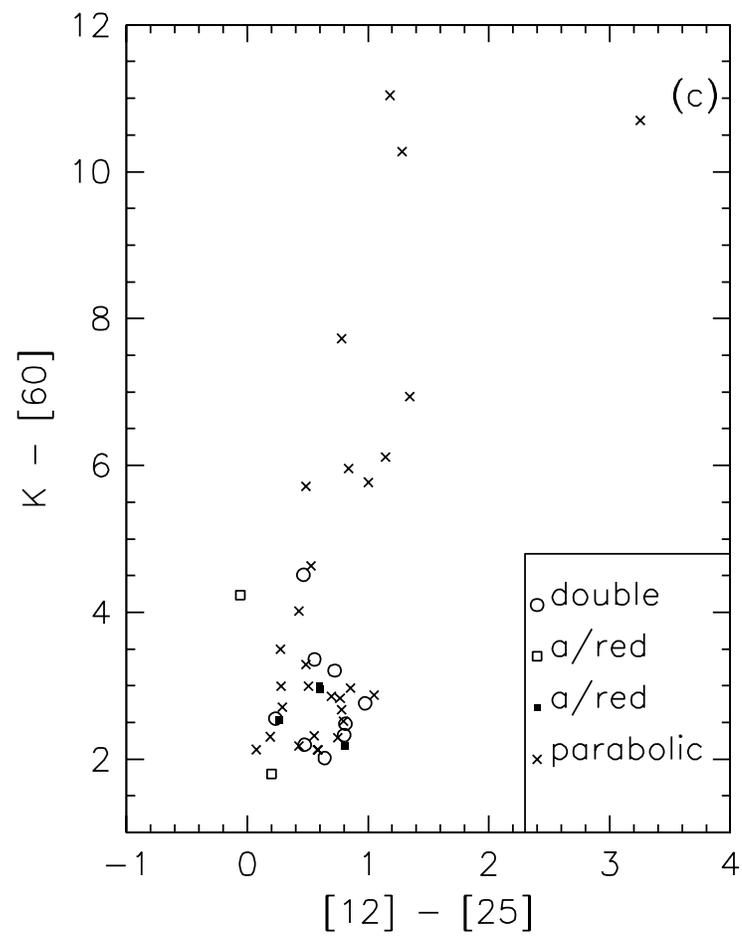
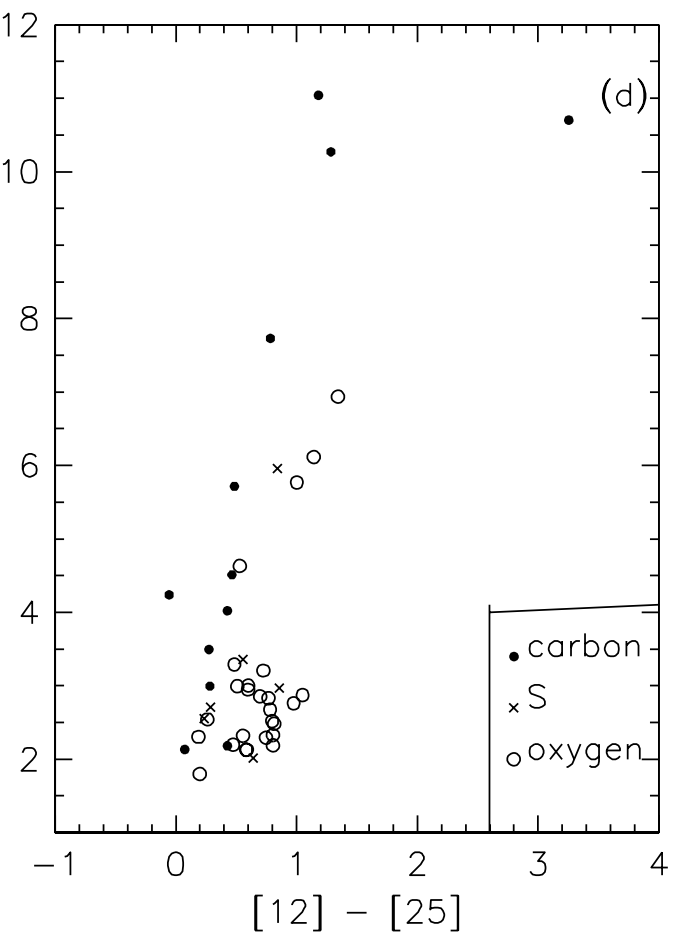
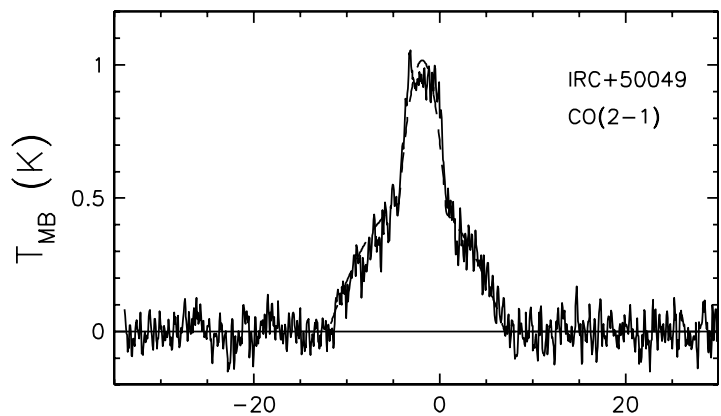
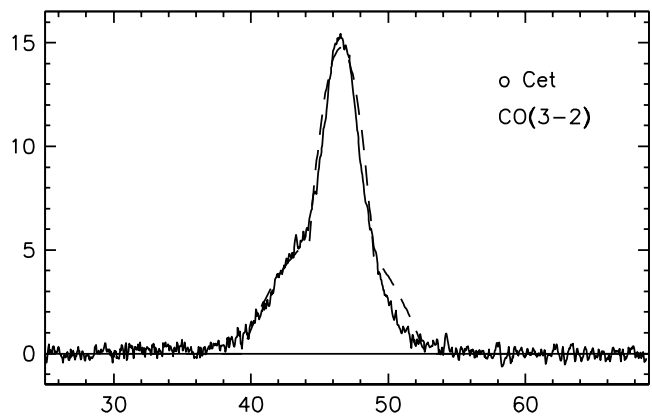
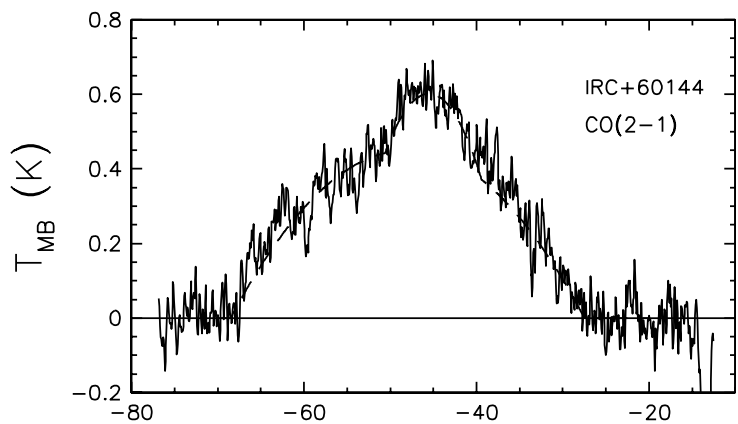
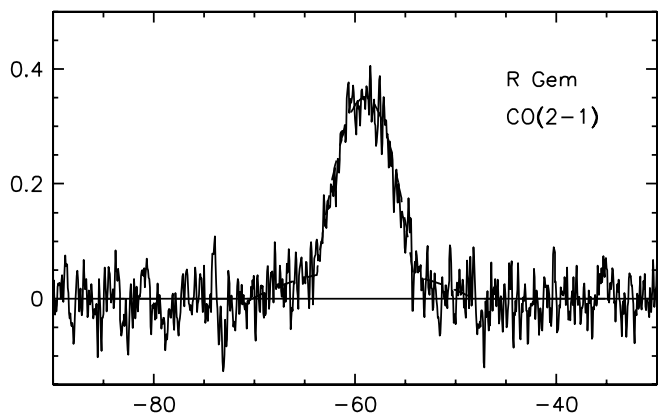
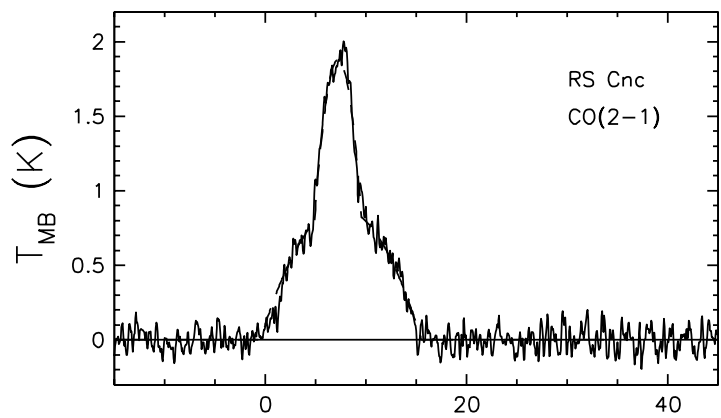
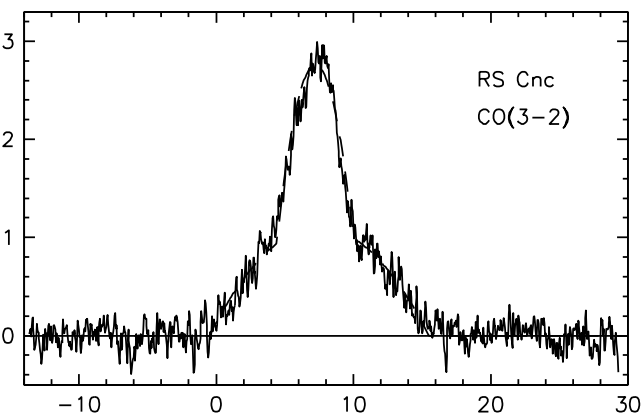
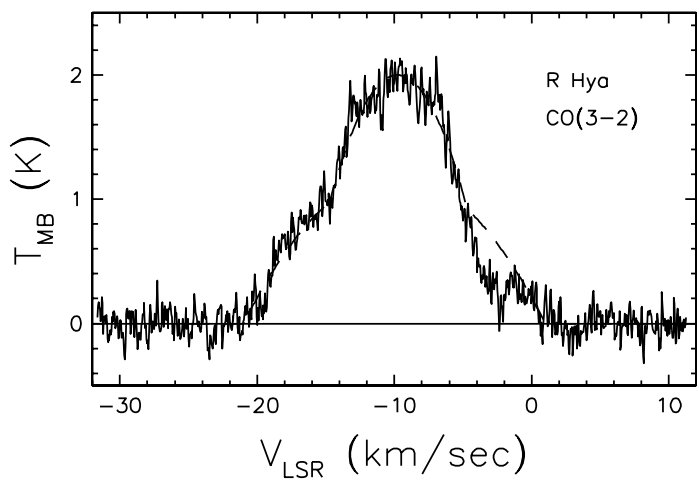
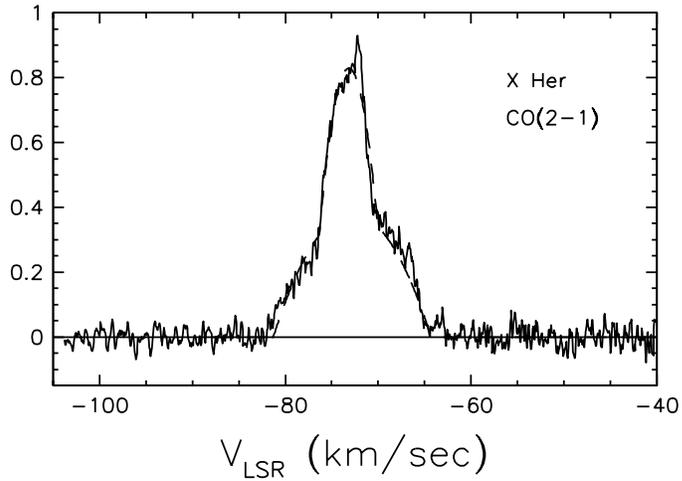
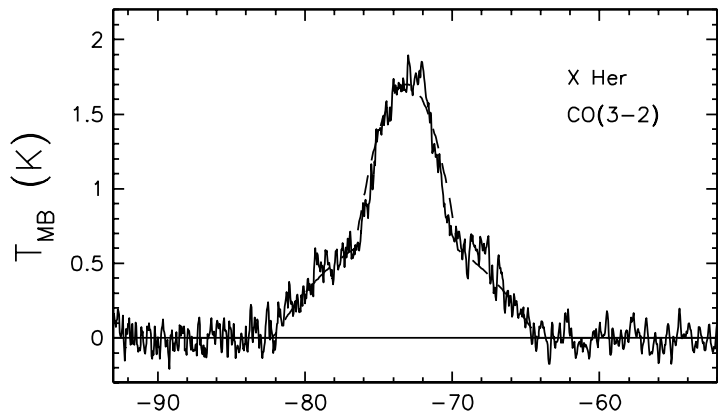
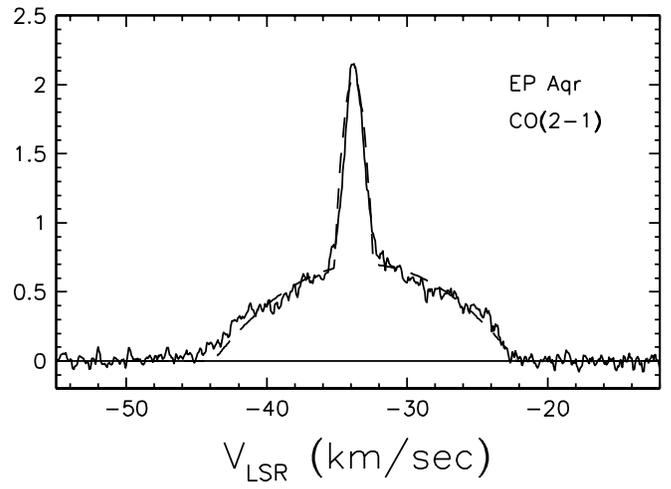
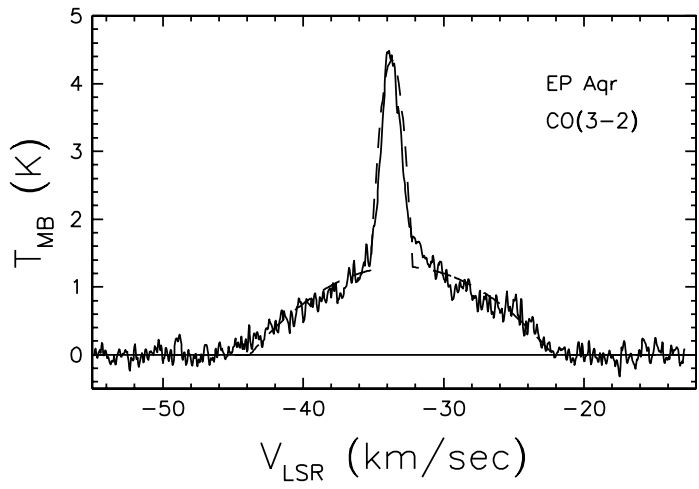
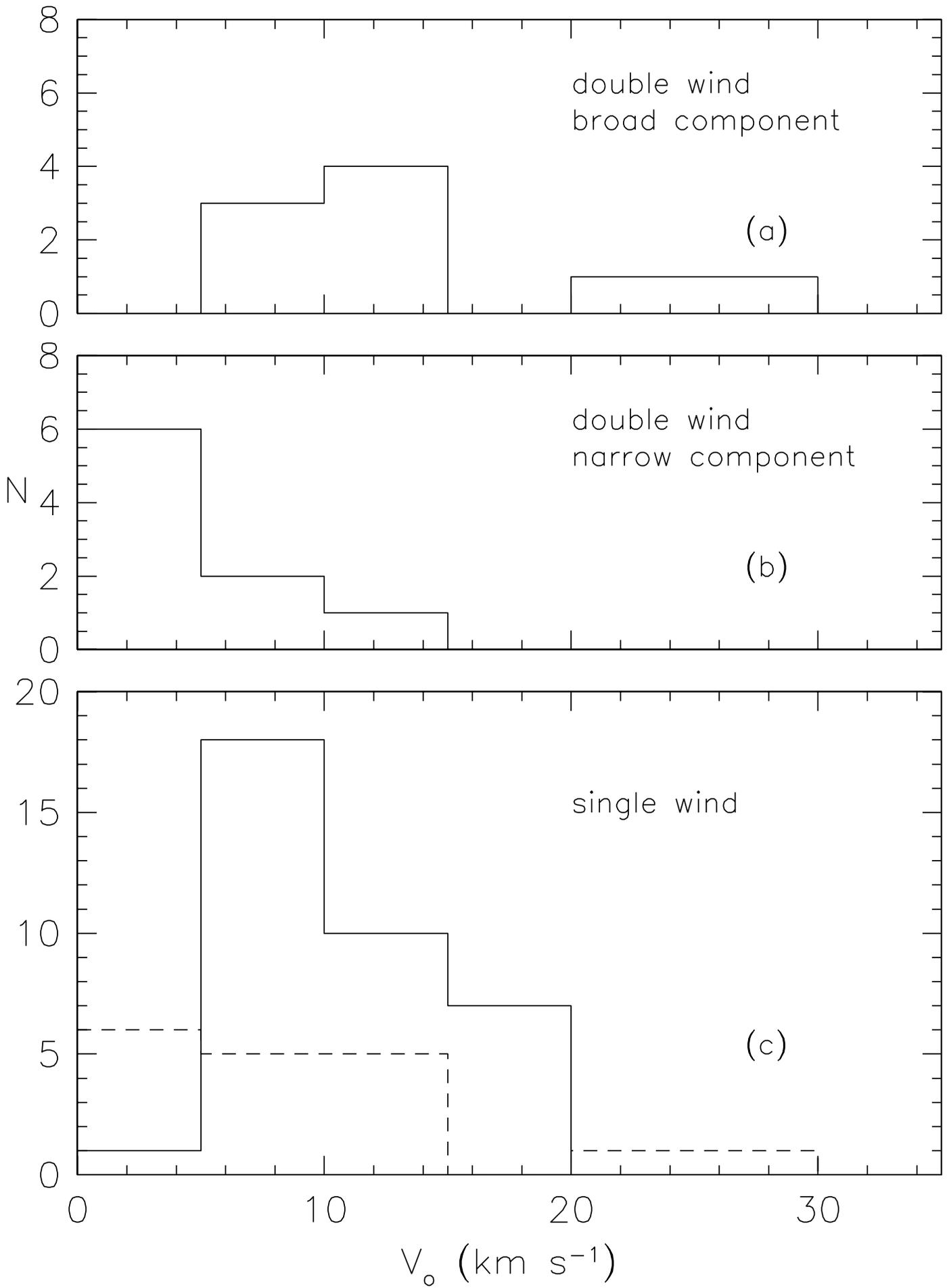
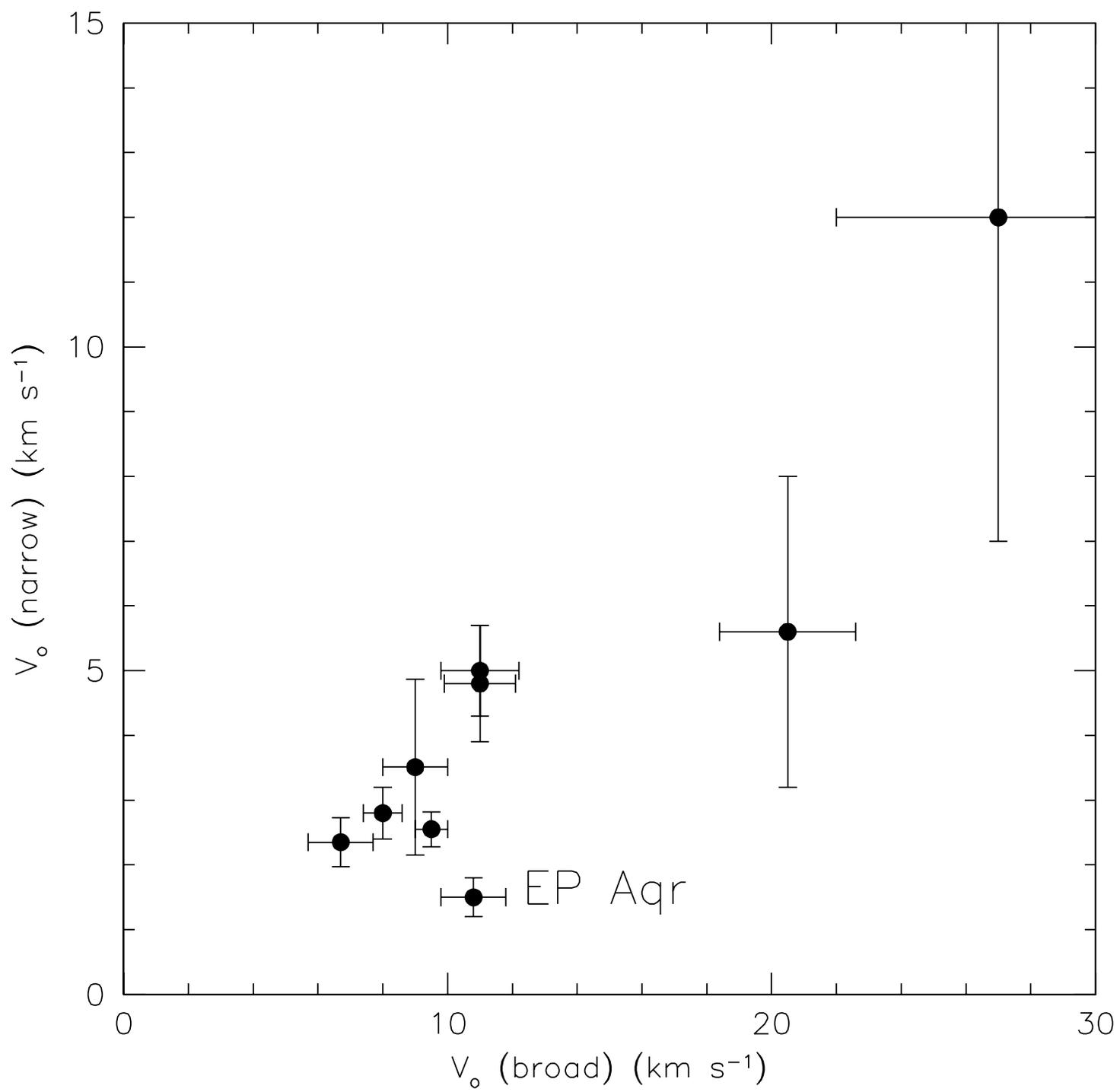
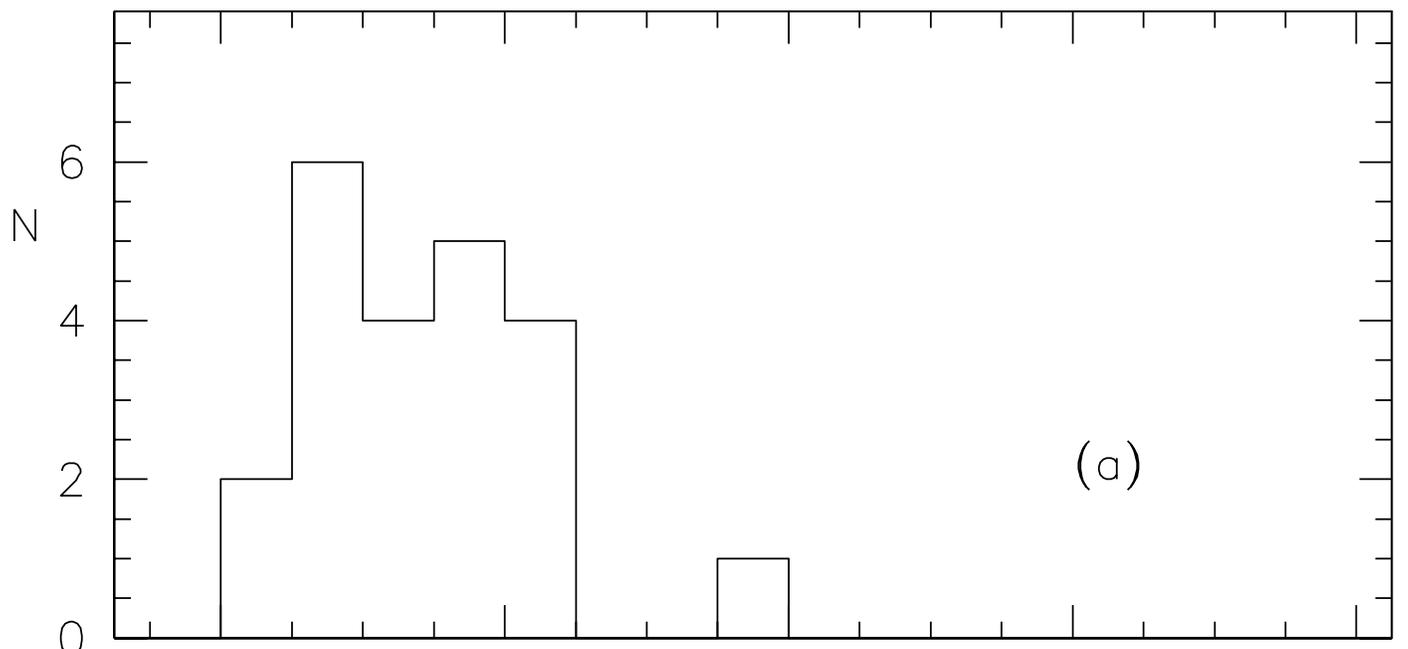
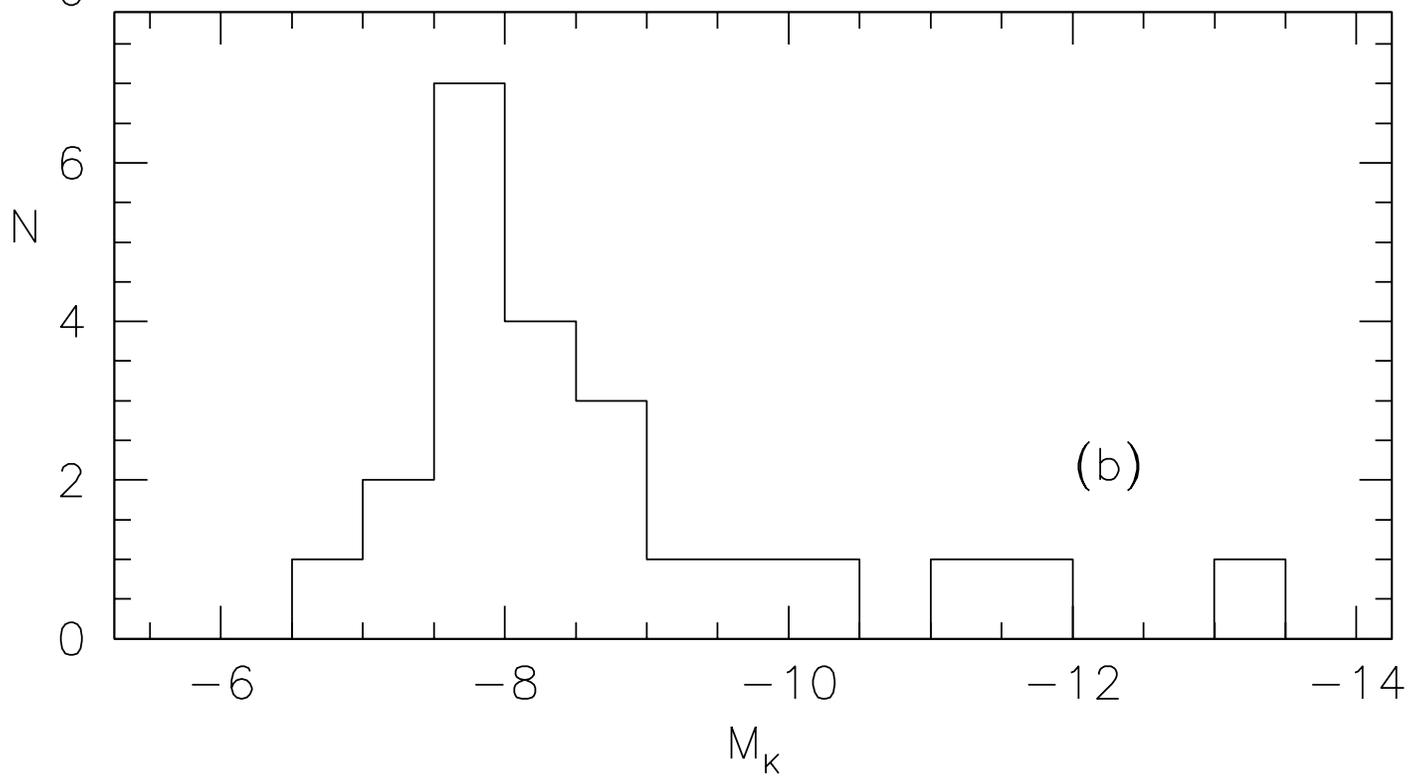
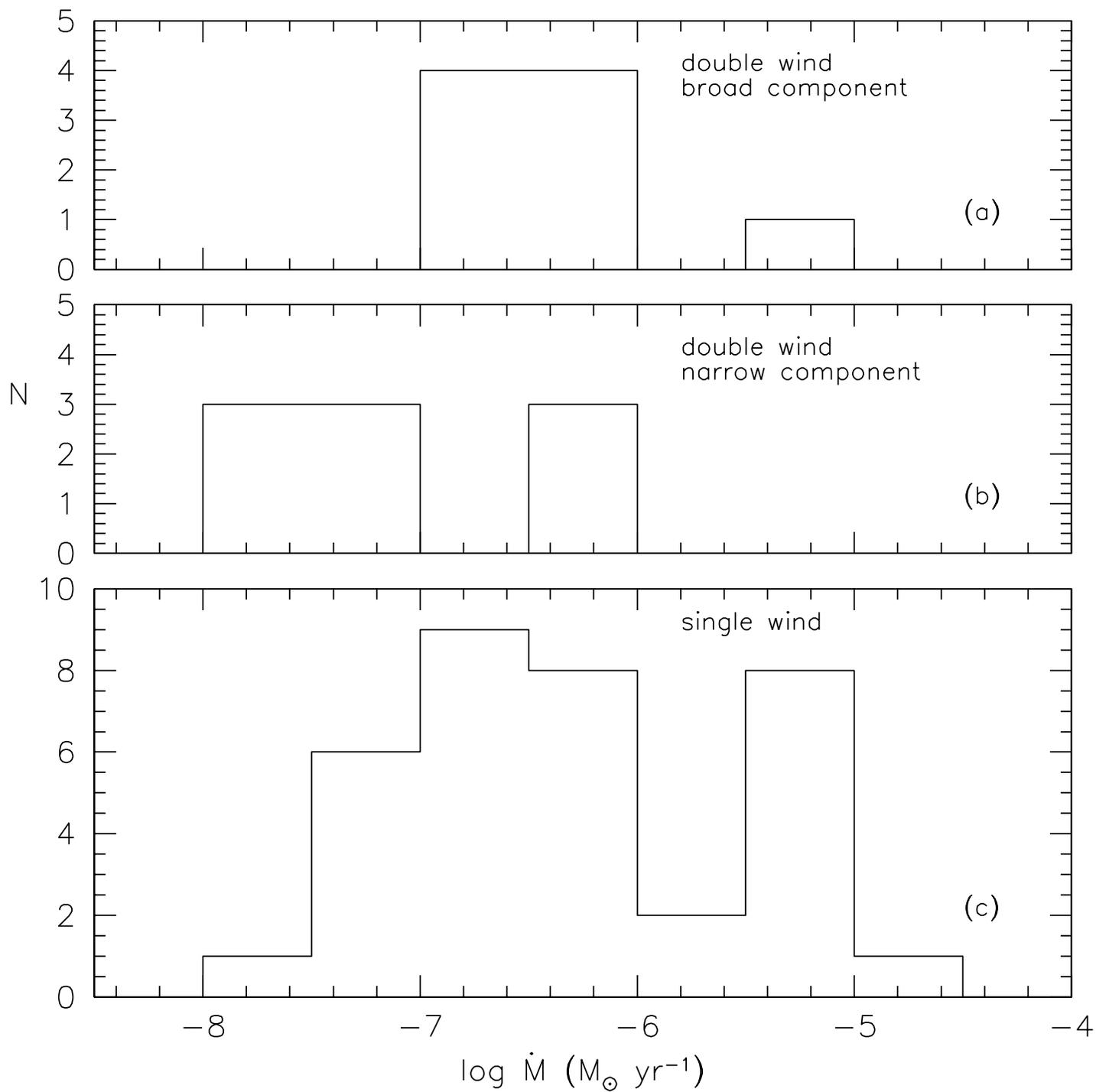
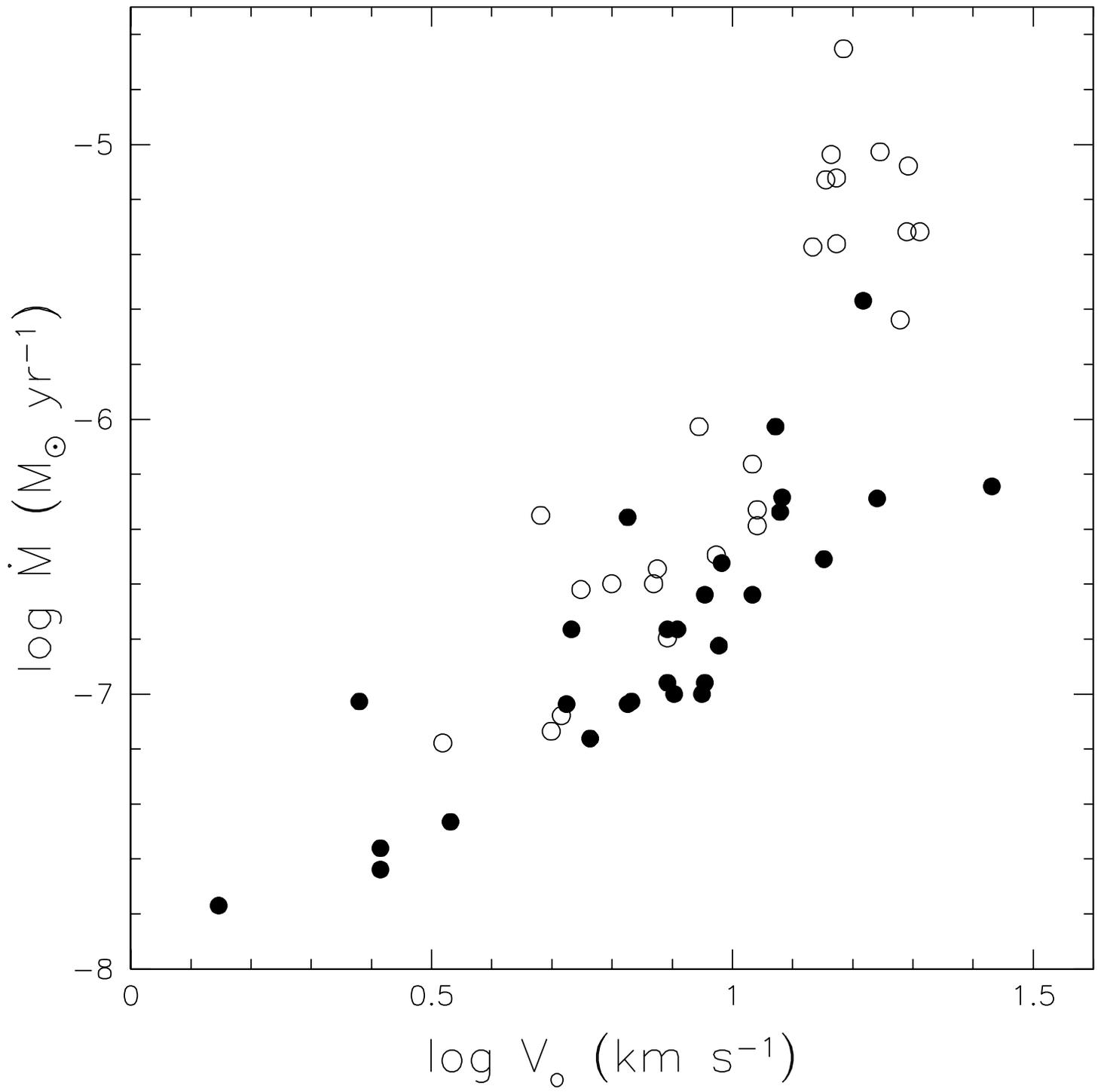
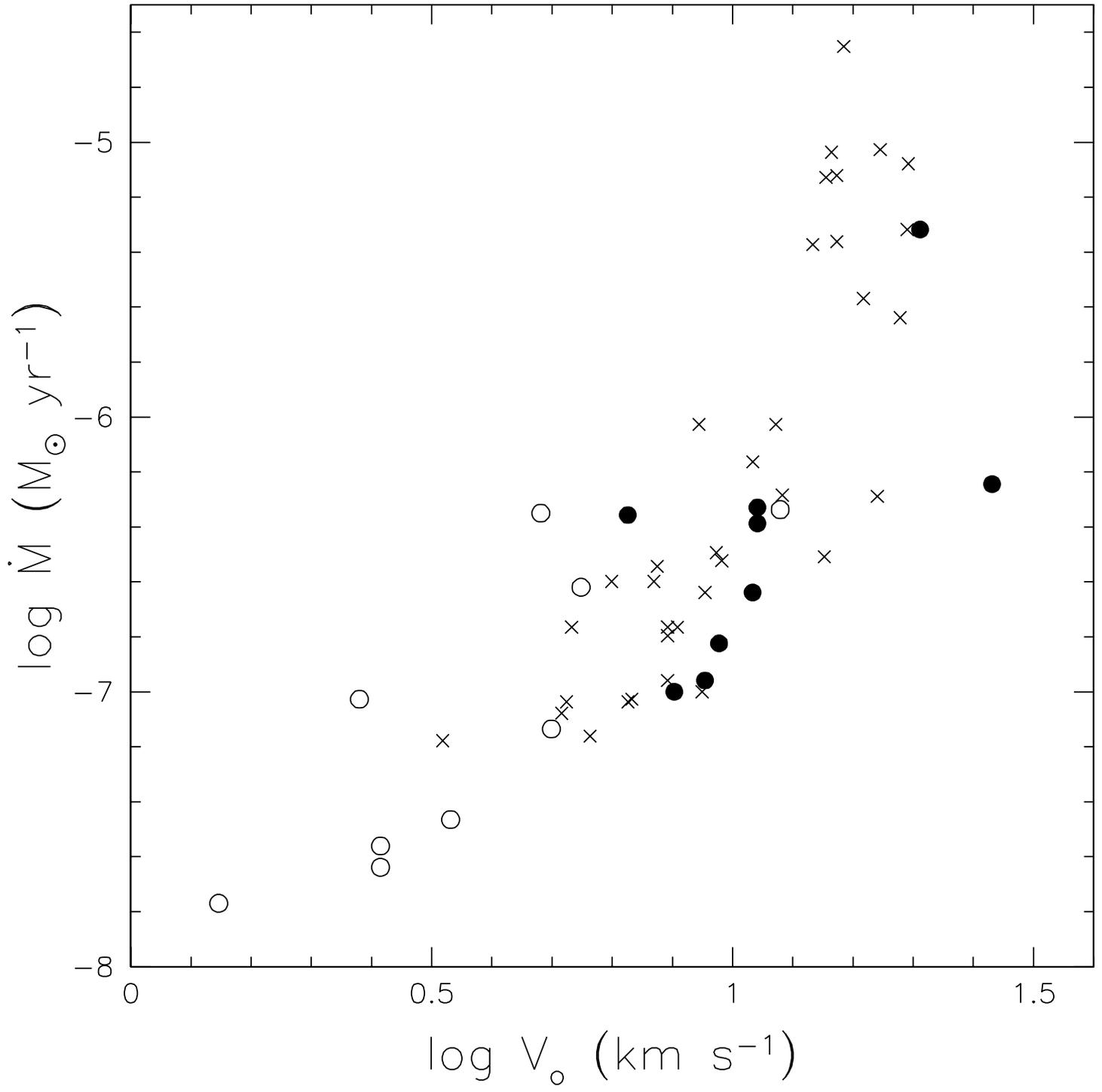
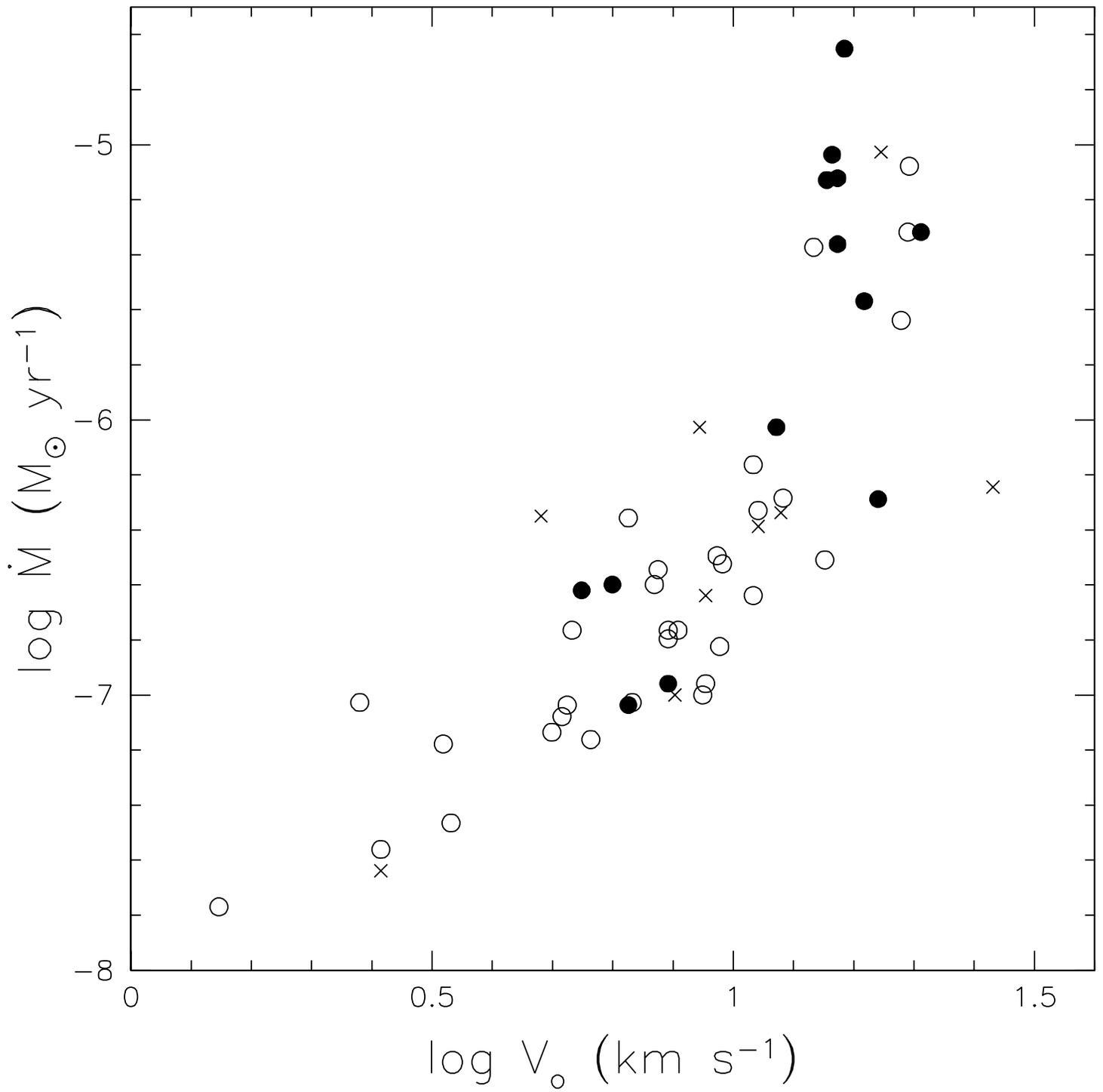
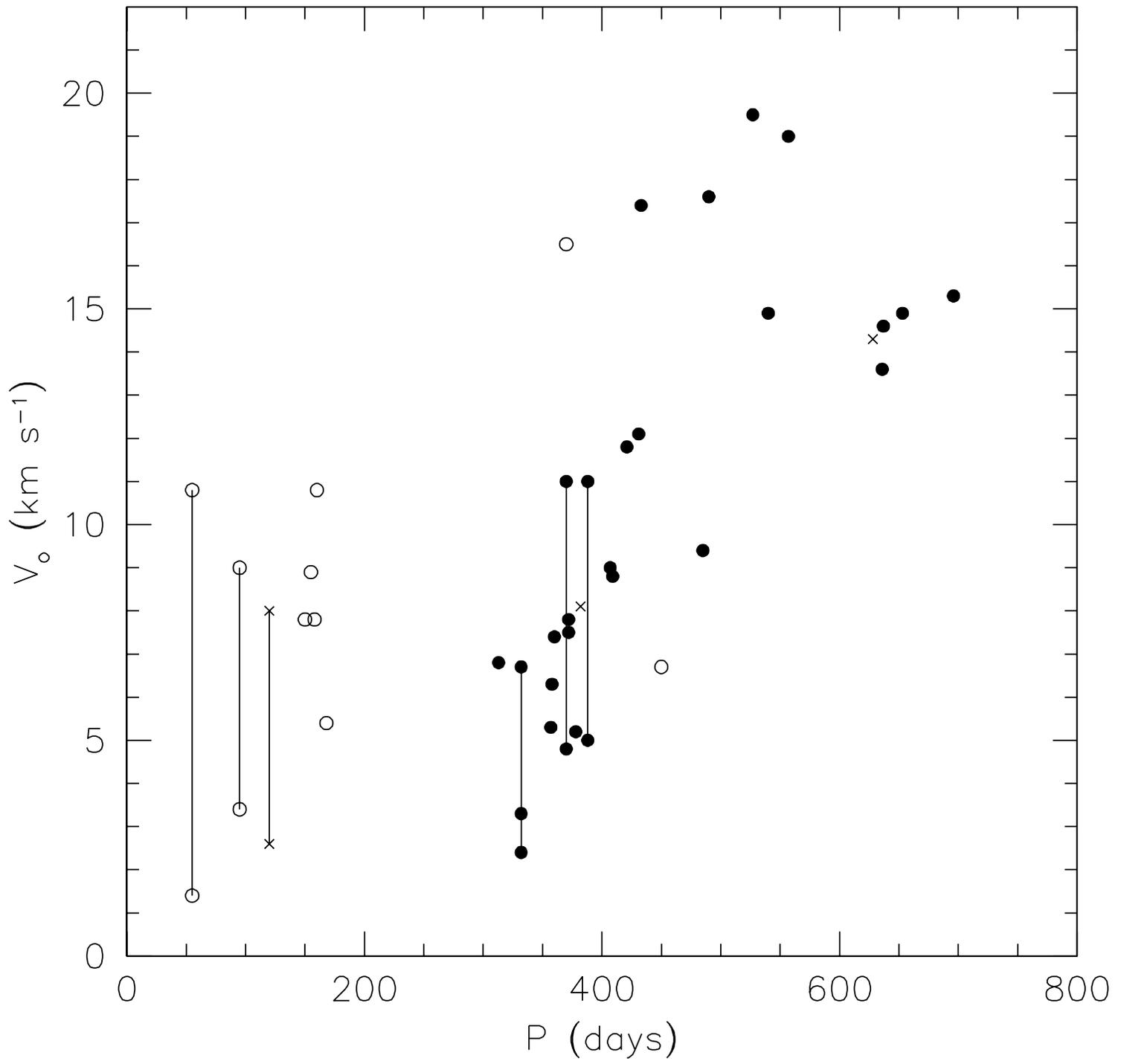
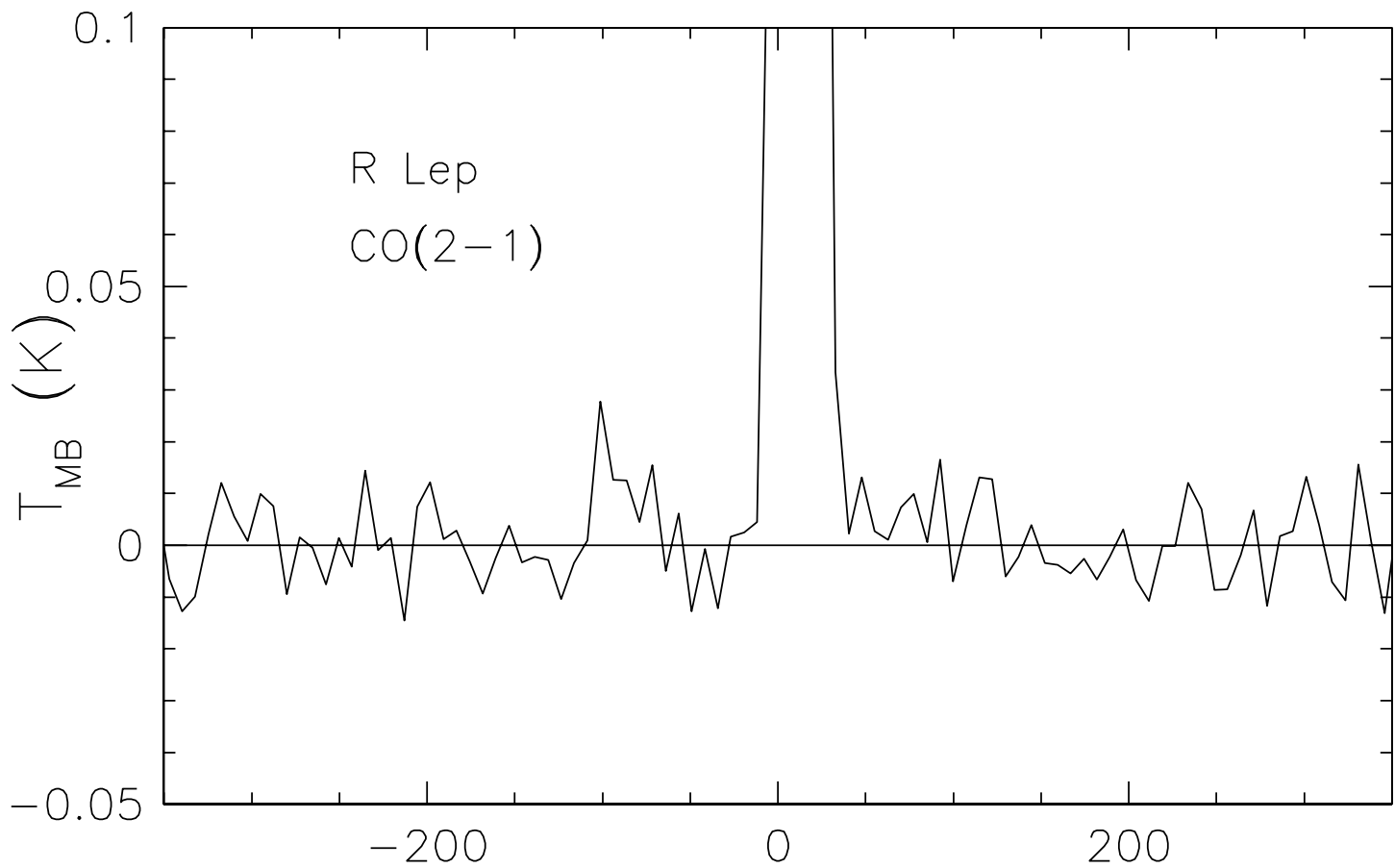
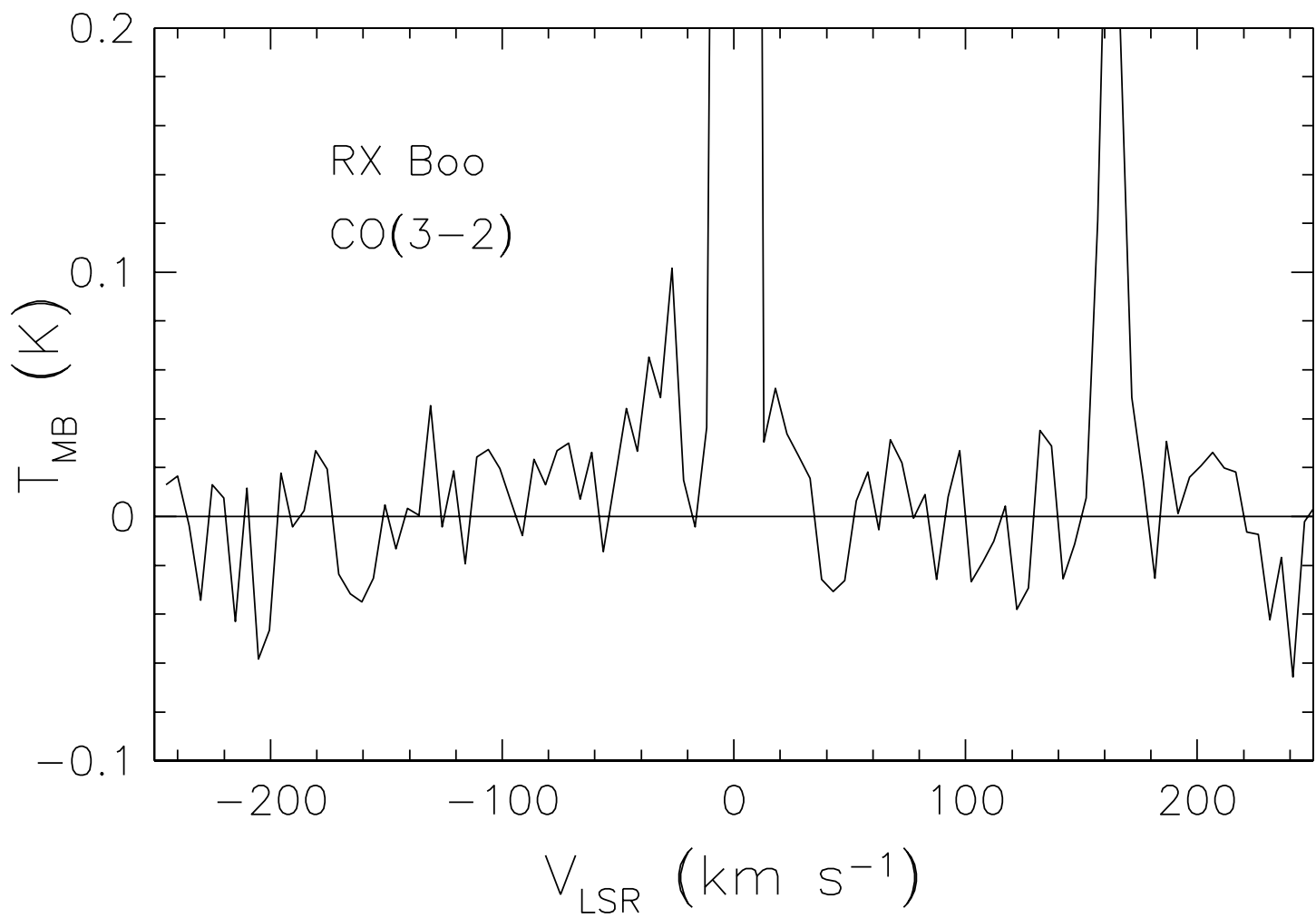
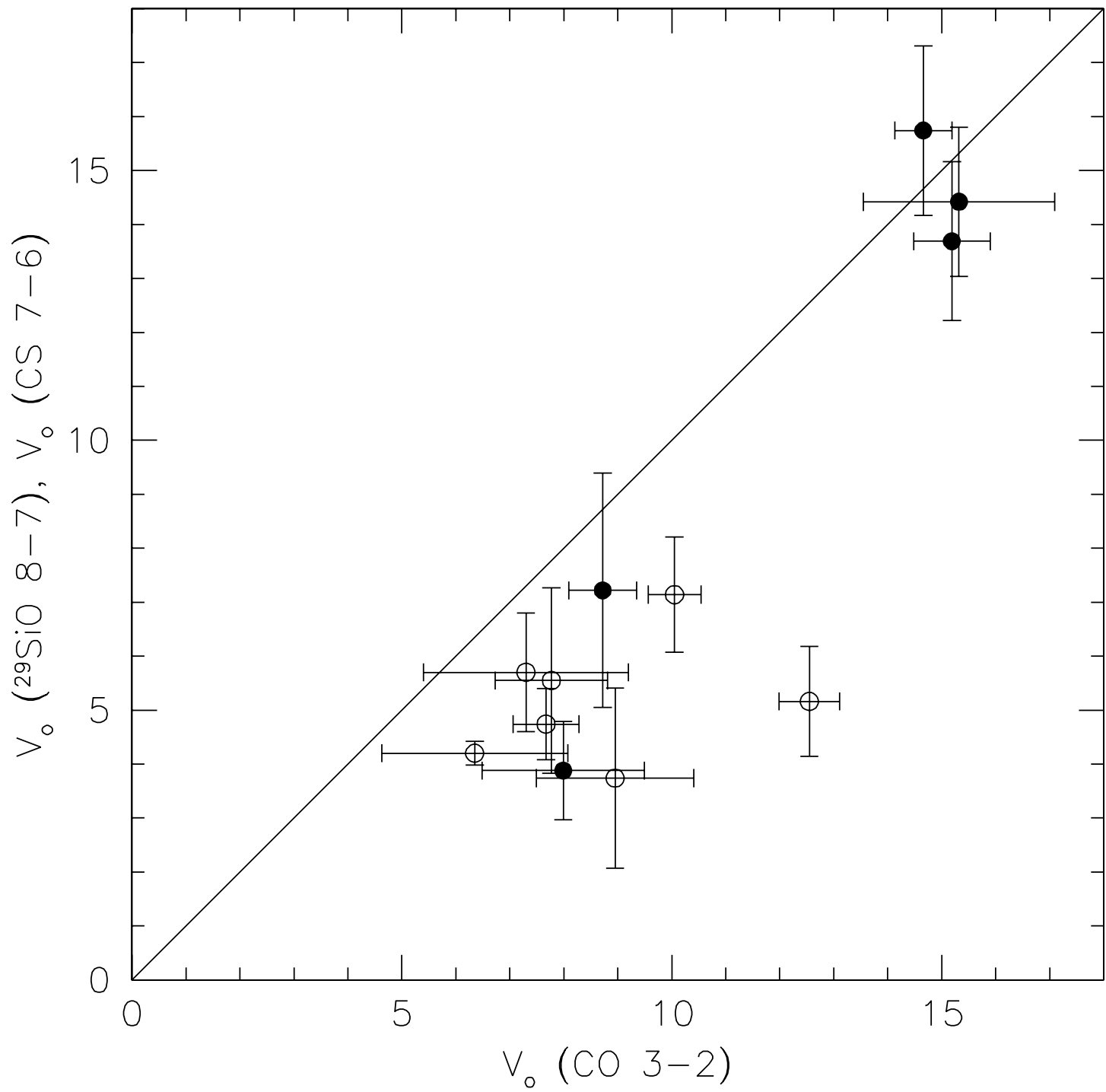
\end